\documentclass{jfm}
\usepackage{graphicx}
\usepackage{epstopdf, epsfig}
\usepackage{natbib}
\usepackage[usenames,dvipsnames]{color}
\usepackage{color}
\usepackage{amssymb}
\usepackage{amsmath}
\usepackage{bm} 
\usepackage{latexsym}
\usepackage[left]{lineno}

\newcommand\Reytau{\mathrm{Re}_{\tau}}
\newcommand\dd{\mathrm{d}}

\shorttitle{locating the MVP overlap}

\shortauthor{P. Monkewitz \& H. Nagib}

\title{The hunt for the K\'arm\'an ``constant'' revisited}

\author{Peter A. Monkewitz\aff{1}
  \corresp{\email{peter.monkewitz@epfl.ch}}
 \and Hassan M. Nagib\aff{2}}

\affiliation{\aff{1}\'Ecole Polytechnique F\'ed\'erale de Lausanne (EPFL), CH-1015, Lausanne, Switzerland
\aff{2}Illinois Institute of Technology, Chicago, IL 60616, USA}

\begin{document}

\maketitle   

\begin{abstract}
The logarithmic law of the wall, joining the inner, near-wall mean velocity profile (abbreviated MVP) in wall-bounded turbulent flows to the outer region, has been a permanent fixture of turbulence research for over hundred years, but there is still no general agreement on the value of the pre-factor, the inverse of the K\'arm\'an ``constant'' $\kappa$, or on its universality.
The choice diagnostic tool to locate logarithmic parts of the MVP is to look for regions where the indicator function $\Xi$ (equal to the wall-normal coordinate $y^+$ times the mean velocity derivative $\dd U^+/\dd y^+$) is constant. In pressure driven flows however, such as channel and pipe flows, $\Xi$ is significantly affected by a term proportional to the wall-normal coordinate, of order $\mathcal{O}(\Reytau^{-1})$ in the inner expansion, but moving up across the overlap to the leading $\mathcal{O}(1)$ in the outer expansion.
Here we show that, due to this linear overlap term, $\Reytau$'s well beyond $10^5$ are required to produce one decade of near constant $\Xi$ in channels and pipes.
The problem is resolved by considering the common part of the inner asymptotic expansion carried to $\mathcal{O}(\Reytau^{-1})$, and the leading order of the outer expansion. This common part contains
a \textit{superposition} of the log law and a linear term $S_0 \,y^+\Reytau^{-1}$, and corresponds to the linear part of $\Xi$, which, in channel and pipe, is concealed up to $y^+ \approx 500-1000$ by terms of the inner expansion. A new and robust method is devised to simultaneously determine $\kappa$ and $S_0$ in pressure driven flows at currently accessible $\Reytau$'s, yielding $\kappa$'s which are consistent with the $\kappa$'s deduced from the Reynolds number dependence of centerline velocities. A comparison with the zero-pressure-gradient turbulent boundary layer, henceforth abbreviated ZPG TBL, further clarifies the issues and improves our understanding.
\end{abstract}

\section{\label{sec1}Introduction}

In the following, the standard non-dimensionalization is adopted, with the ``inner'' or viscous length scale $\widehat{\ell} \equiv (\widehat{\nu}/\widehat{u}_\tau)$, where $\widehat{u}_\tau \equiv (\widehat{\tau}_w/\widehat{\rho})^{1/2}$, $\widehat{\rho}$ and $\widehat{\nu}$ are the friction velocity, density and dynamic viscosity, respectively, with hats denoting dimensional quantities. The resulting non-dimensional mean velocity is $U^+ \equiv (\widehat{U}/\widehat{u}_\tau)$, and the
inner and outer wall-normal coordinates are $y^+=\widehat{y}/\widehat{\ell}$ and $Y=y^+/\Reytau$, respectively, with $\Reytau\equiv \widehat{\mathcal{L}}/\widehat{\ell}$ the friction Reynolds number and $\widehat{\mathcal{L}}$ the outer length scale such as channel half width, pipe radius or boundary layer thickness.
\subsection*{The log-law and matched asymptotic expansions (abbreviated MAE)}

The logarithmic law for the mean velocity in wall-bounded turbulent flows goes back to the celebrated work of \citet{vonKarman34} and \citet{Millikan} and is firmly rooted in the framework of MAE \citep[see e.g.][]{KC85,WilcoxP,Panton2005}, where it represents a key term of the overlap layer between the inner and outer mean velocity expansions, $U^+_{\mathrm{in}}(y^+)$ and $U^+_{\mathrm{out}}(Y)$. Its traditional form is $\kappa^{-1} \ln{y^+}$, with $\kappa$ the K\'arm\'an ``constant'', or rather parameter, as its flow dependence is confirmed by the present work.

By definition, this logarithm is, within the overlap, common to both the leading order inner and outer expansions, where it takes the form $\kappa^{-1} [\ln{Y} + \ln{\Reytau}]$. Hence, $\kappa$ can be equally well determined from $U^+(y^+)$ or $U^+(Y)$, but the choice is not as trivial as it seems. The matching also involves some subtleties, which may not be so well known:\newline

\begin{enumerate}
\item For the asymptotic matching of inner and outer expansions, the term $\kappa^{-1} \ln{\Reytau}$ in the outer expansion has to be treated as an $\mathcal{O}(1)$ term, according to the principle of ``block matching'', which has been introduced in MAE by \citet{CL73} to handle terms containing powers of the logarithm of the small parameter $\epsilon$. Thereby, all the terms proportional to $\epsilon^n(\ln{\epsilon})^m$ are regrouped into the same ``block order'' $n$, and have to be treated simultaneously for the matching. This general concept was developed to treat MAE problems in 2D acoustics, where logarithms and powers of logarithms abound and $\epsilon$ is typically the ratio of acoustic wavelength to distance from the source.
    This concept has actually been used for a long time by the turbulent boundary layer community without being formalized. To match inner and outer expansions of the MVP across the overlap, $\kappa^{-1}\ln{y^+}$ in the inner expansion has always been identified with $\kappa^{-1}\ln{Y} + \kappa^{-1}\ln{\Reytau}$ in the outer expansion, where $\ln{\Reytau}$ has been treated as an $\mathcal{O}(1)$ term. Obviously, there is no match in the inner expansion for $\kappa^{-1}\ln{\Reytau}$ alone.
\item Furthermore, if the outer expansion is of the well accepted form $\kappa^{-1} [\ln{Y} + \ln{\Reytau}]$ plus an $\mathcal{O}(1)$ function of $Y$ \citep[see e.g.][]{Coles56}, plus terms of order $\mathcal{O}(\Reytau^{-1})$ and higher, the leading order centerline velocity in channels and pipes is $\kappa^{-1} \ln{\Reytau}$ plus a constant, as discussed in \citet{NagibTSFP10} and \citet{Monk21}, for instance. This equality of overlap and centerline $\kappa$ could only be relaxed, if the outer expansion contained an additional $\mathcal{O}(1)$ term proportional to $[\exp{(-\mathrm{const.}/Y)}\,\ln{\Reytau}]$, which becomes transcendentally small for $Y \to 0$ \citep[see e.g.][for a discussion of transcendentally small terms]{WilcoxP}. Lacking any evidence for such a term, the $\kappa$'s extracted from overlap profiles and from the $\Reytau$ dependence of the centerline velocity \textit{must be identical} !
\end{enumerate}

\subsection*{The role of the additional linear term in the channel and pipe MVP overlap}

Traditionally, the MVP overlap in all wall-bounded turbulent flows has been associated with a purely logarithmic region, readily identified with the log-indicator function
\begin{equation}
\Xi \equiv y^+\,\frac{\dd U^+}{\dd y^+} \equiv Y\,\frac{\dd U^+}{\dd Y} \quad ,
\label{Xi}
\end{equation}
which is constant whenever $U^+$ is a linear function of $\ln{y^+}$. It is noted, however, that an interval of constant $\Xi$ is not automatically an inner-outer overlap, as there are additional requirements in MAE. Specifically, the center of the overlap has to scale on the intermediate variable $(y^+\,Y)^{1/2}$, and its extent has to expand with $\Reytau$.
In technical MAE terms, looking for a region of constant $\Xi$, i.e. a simple log-law, amounts to consider the basic (1$\mathcal{O}$inner/1$\mathcal{O}$outer) common part or overlap.
Here and in the following, ``(n$\mathcal{O}$inner/m$\mathcal{O}$outer) overlap'', is a shorthand for an overlap constructed from an inner asymptotic expansion of n-th order and its outer counterpart of m-th order.

The problems with this traditional approach
stem from additional terms in the overlap region, discussed early on by \citet{Yajnik70} and \citet{AfzalY73}, among others. Of particular relevance is the linear term $S_0\,y^+/\Reytau$ in the $U^+$ overlap profile of channels and pipes, which represents an $\mathcal{O}(\Reytau^{-1})$ correction of the \textit{inner-scaled} indicator function $\Xi(y^+)$, as discussed for instance by \citet{Jimenez07}, \citet[][sec. 3.1]{LM15} and \citet{Luchini17}, who has argued that the coefficient $S_0$ of this linear term $S_0\,y^+/\Reytau$ is proportional to the pressure gradient (see appendix \ref{appC} for a review of this issue). As this linear term is of higher order in the inner expansion, it is not included in the (1$\mathcal{O}$inner/1$\mathcal{O}$outer) MVP overlap, despite moving up to $\mathcal{O}(1)$ in the \textit{outer-scaled} indicator function $\Xi(Y)$.
This follows formally from the overlap description in terms of the intermediate variable $\eta = y^+\Reytau^{-1/2} = Y\,\Reytau^{+1/2}$, where the linear term $S_0\,y^+/\Reytau \equiv S_0\,Y$ is of order $\mathcal{O}(\Reytau^{-1/2})$ relative to the $\mathcal{O}(1)$ log-law.

However, at the Reynolds numbers where data are available, the basic (1$\mathcal{O}$inner/1$\mathcal{O}$outer) overlap to determine $\kappa$ is not very helpful in the presence of an additional linear overlap term $S_0\,y^+/\Reytau$, such as in channels and pipes, where the overlap indicator function takes the form
\begin{equation}
\Xi_{\mathrm{OL}} = \frac{1}{\kappa} + \frac{S_0\,y^+}{\Reytau} + \mathrm{H.O.T.} \equiv \frac{1}{\kappa} + S_0\,Y  + \mathrm{H.O.T.}\quad ,
\label{XiOL}
\end{equation}
and H.O.T. designates higher order linear terms considered only for the channel in section \ref{sec2}.

\begin{figure}
\center
\includegraphics[width=0.65\textwidth]{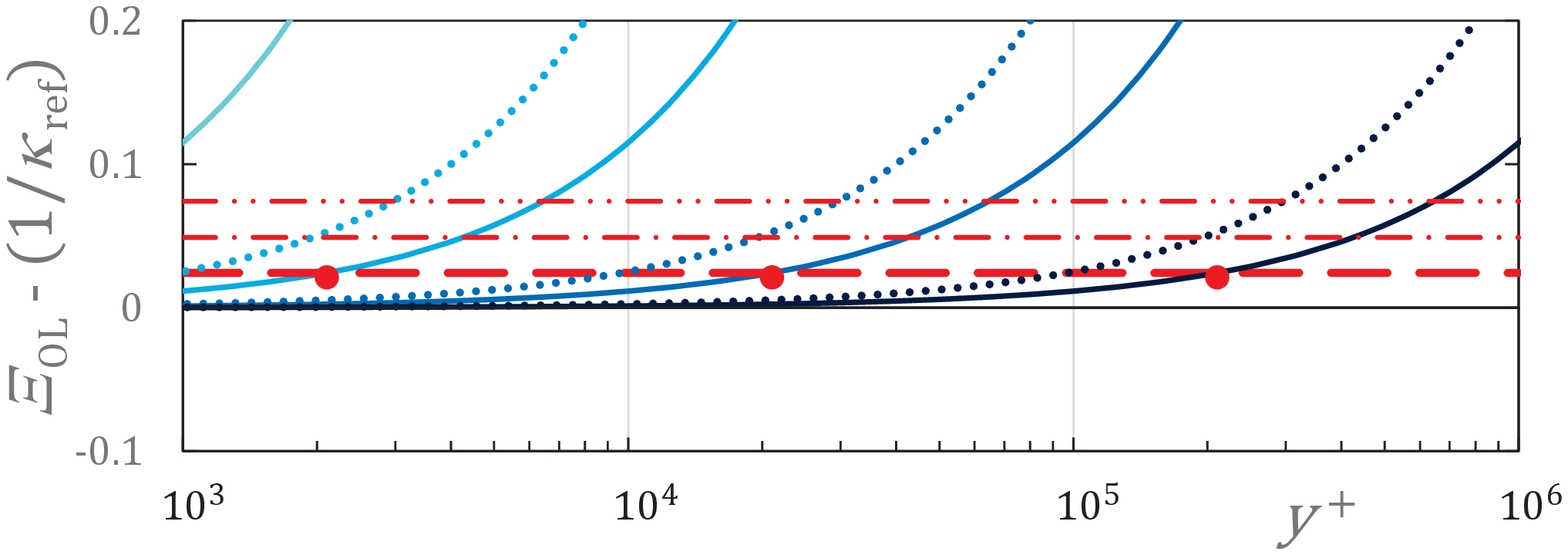}
\caption{\label{refFig1}
Illustration of the effect of the linear term $S_0\, y^+/\Reytau$  on the estimate of $\kappa^{-1}$ from the overlap $\Xi_{\mathrm{OL}}$ (equation \ref{XiOL}). (red) $- - -$, $-\cdot -\cdot$, $-\cdot\cdot\,-$, $\kappa$ errors of -1\%, -2\% and -3\% relative to $\kappa_{\mathrm{ref}}=0.417$.
Linear deviations $S_0\, y^+/\Reytau$ from the baseline log-law for $\Reytau = 10^4,\, 10^5,\, 10^6$ and $10^7$ (increasingly dark blue) with $S_0=1.15$ (---) and $S_0=2.5$ ($\cdot\cdot\cdot$).
(red) $\bullet$, locations where the linear terms with $S_0 = 1.15$ induce a -1\% error of $\kappa$. }
\end{figure}

Ignoring the linear contribution to the overlap $\Xi_{\mathrm{OL}}$ in equation (\ref{XiOL}) has been the main reason for the lack of agreement on $\kappa$ values. The example of figure \ref{refFig1} shows, that determining $\kappa$ with an error below 1\% from a
region of sufficiently constant $\Xi_{\mathrm{OL}}$, extending from, say, $y^+ = 10^3$ to $10^4$, requires a very large Reynolds number of $\Reytau \approx (10^6\,\kappa\,S_0)$.
Note, that the reason for considering only the region of $y^+\geq 10^3$ in figure \ref{refFig1} is the ``hump'' or ``bulge'' of $\Xi$ below $y^+\approx 10^3$ on top of the linear overlap (\ref{XiOL}), which will be discussed in the next section \ref{sec2}.

The problem of the non-negligible linear term in the overlap of channels and pipes is resolved by moving to the (2$\mathcal{O}$inner/1$\mathcal{O}$outer) overlap, which includes the linear term $S_0\,y^+/\Reytau \equiv S_0\,Y$, because it is present in \textit{both} the limit $y^+ \to \infty$ of the inner expansion carried to the order $\mathcal{O}(\Reytau^{-1})$, $U^+_{\mathrm{in}}(y^+\gg 1) \sim \kappa^{-1}\ln{y^+} + B_0 + B_1/\Reytau + S_0\,y^+/\Reytau$ (see equation \ref{Uout}), and in the limit $Y \to 0$ of the leading order outer expansion
$U^+_{\mathrm{out}}(Y\ll 1) \sim \kappa^{-1}[\ln{Y} + \ln{\Reytau}] + B_0 + S_0\,Y$. This (2$\mathcal{O}$inner/1$\mathcal{O}$outer) overlap corresponds to the \textit{linear} outer-scaled indicator function $\Xi_{\mathrm{OL}} = \kappa^{-1} + S_0\,Y$  (equation \ref{XiOL}), which, up to $y^+ \approx 10^3$, is ``buried'' under an inner-scaled ``hump'' or ``bulge'', further discussed in section \ref{sec2} and clearly visible in figure \ref{refFig3} for channel DNS beyond a $\Reytau$ of around 4,000 and, less pronounced, in figure \ref{refFig5} for pipe DNS.
This supports the conclusion of \citet{Monk21} about the ``late start'' of the channel overlap.

Beyond $y^+ \approx 10^3$, the linear overlap $[\kappa^{-1} + S_0\,Y]$ of $\Xi_{\mathrm{OL}}$ becomes clearly visible in these figures and is seen to extend to $Y \approx 0.4-0.5\,$.

From the above outer-scaled form (\ref{XiOL}) of the overlap $\Xi_{\mathrm{OL}}$ it follows conclusively, that the ``humps'' of $\Xi$ on top of the linear overlap, seen in all the channel and pipe profiles below a $y^+$ of roughly $10^3$, belong to the \textit{inner} expansion. Consequently, the short horizontal or near-horizontal parts of $\Xi$ within these ``humps'', seen for instance in \citet[][fig. 3]{LM15},  as well as in figures \ref{refFig3} and \ref{refFig5} of the present paper, are not overlap log laws, but locally logarithmic or nearly logarithmic regions of the inner expansion.

The interpretation of the indicator function $\Xi$ is further complicated by a surprisingly large variability between different channel DNS, and even more so between pipe DNS. This variability has different sources, such as domain size, computational scheme, convergence of computation and grid spacing. The particular effect of grid spacing in the outer flow region is highlighted in appendix \ref{appB}. The analysis suggests that there is a critical grid spacing $\Delta y^+$ of 3 to 4, which, when exceeded towards the centerline, leads to a decrease of the effective $\Reytau$ in the central flow region (see figures \ref{refFig3} and \ref{refFig12} of appendix \ref{appB}).
\subsection*{Outline of the paper}
The purpose of this paper is to clarify both the location, extent and the functional form of the inner-outer overlap in channels and pipes, and to propose a novel robust method to extract $\kappa$ from MVPs in pressure driven flows. A comparison with the ZPG TBL further clarifies the issues. The paper is organized as follows:

\begin{enumerate}
\item In section \ref{sec2}, an improved outer fit of the mean velocity derivative in channels is developed from DNS, with additional details provided in appendix \ref{appA}. The resulting outer fit of the indicator function $\Xi$ is compared to different channel DNS and the variability of the results is correlated with the different choices of computational grid spacing.
    The superposition of log law and linear term in the overlap is supported by the experimental data of \citet{ZDN03} and \citet{SchultzFlack2013} obtained in channels of aspect ratio $\approx 8$. The simultaneous determination of the two overlap parameters $\kappa$ and $S_0$ is performed with a new robust method, presented in figure \ref{refFig4} and believed more discriminating than the iterative method of \citet{Luchini18}.
\item Section \ref{sec3} then presents an analysis of three pipe flow DNS by \citet{ElKhoury2013}, \citet{pirozzoli_pipe2021} and \citet{Yao2023}, which show considerable differences of $\Xi(Y)$. On the other hand, the $\Xi$ for the Superpipe data of \citet{ZS98}, \citet{McK_thesis} and \citet{Pitot13} are found to be very consistent, and the new method for the determination of overlap parameters yields $\kappa = 0.433$ and $S_0 = 2.5$ for the coefficient of the linear term.
\item In the brief section \ref{sec4}, the findings for channel and pipe flow are contrasted with the ZPG TBL. The experimental data from three independent sources reveal that the TBL indicator function also features a linear part of a significantly higher slope than in channels and pipes, with the crucial difference that this linear part only \textit{starts in the outer region} at $Y=0.11$, and therefore does \textit{not} belong to the overlap.
\item The final section \ref{sec5} summarizes the main results and closes with observations on the universality, or rather non-universality, of the so-called canonical turbulent flows: channel flow, pipe flow and ZPG TBL.
\end{enumerate}

Three appendices complete the paper: Appendix \ref{appA} provides more details on the fit of the channel MVP, used in section \ref{sec2}. Appendix \ref{appB} discusses the likely effect of grid spacing on the results of channel and pipe DNS.
Appendix \ref{appC}, finally reviews different approaches, including the one by \citet{Luchini17}, to take into account the effect of pressure gradient on MVP overlap profiles.

\section{\label{sec2}The outer expansion and the overlap of the indicator function in channels}

To prepare for the analysis of the channel indicator function, the outer mean velocity fit of \citet[][equ. 3.6]{Monk21} is improved and simplified, while maintaining its basic ingredients.
The differences to \citet{Monk21} are that the fitting is started with the mean velocity \textit{derivative}, the $\kappa$ is slightly modified to 0.417,
and the $\mathcal{O}(\Reytau^{-1})$ contribution to $\dd U^+/\dd Y$ is simplified :
\begin{eqnarray}
& \frac{\dd U^+}{\dd Y}\arrowvert_{\mathrm{channel\,out}} = \frac{1}{\kappa\,Y} + \left[S_0 + \frac{1}{\Reytau}\,S_1\right] - \left[\frac{1}{\kappa} + S_0 + \frac{1}{\Reytau}\,S_1\right]\, \frac{\dd W}{\dd Y} \label{UY} \\
& \mathrm{with} \nonumber \\
& \kappa = 0.417, \, S_0 = 1.15, \, S_1 = 380 \quad \mathrm{and} \quad
 \frac{\dd W}{\dd Y} = \frac{\ln{\{\exp{[11\,(Y-0.73)]} +
1\}}}{\ln{\{\exp{[11\,(1-0.73)]} + 1 \}}} \label{UYcofs} \\
& U^+\arrowvert_{\mathrm{channel\,out}} = \frac{\ln{Y}}{\kappa} + \frac{\ln{\Reytau}}{\kappa} +
\left[S_0 + \frac{S_1}{\Reytau}\right]\,Y + \left[B_0 + \frac{B_1}{\Reytau}\right]  \nonumber \\
&  - \left[\frac{1}{\kappa} + S_0 + \frac{S_1}{\Reytau}\right]\,W(Y) \label{Uout} \\
& \mathrm{with}\quad B_0 = 5.45, B_1 = -250 \label{Uoutcofs}
\end{eqnarray}

\begin{figure}
\center
\includegraphics[width=0.65\textwidth]{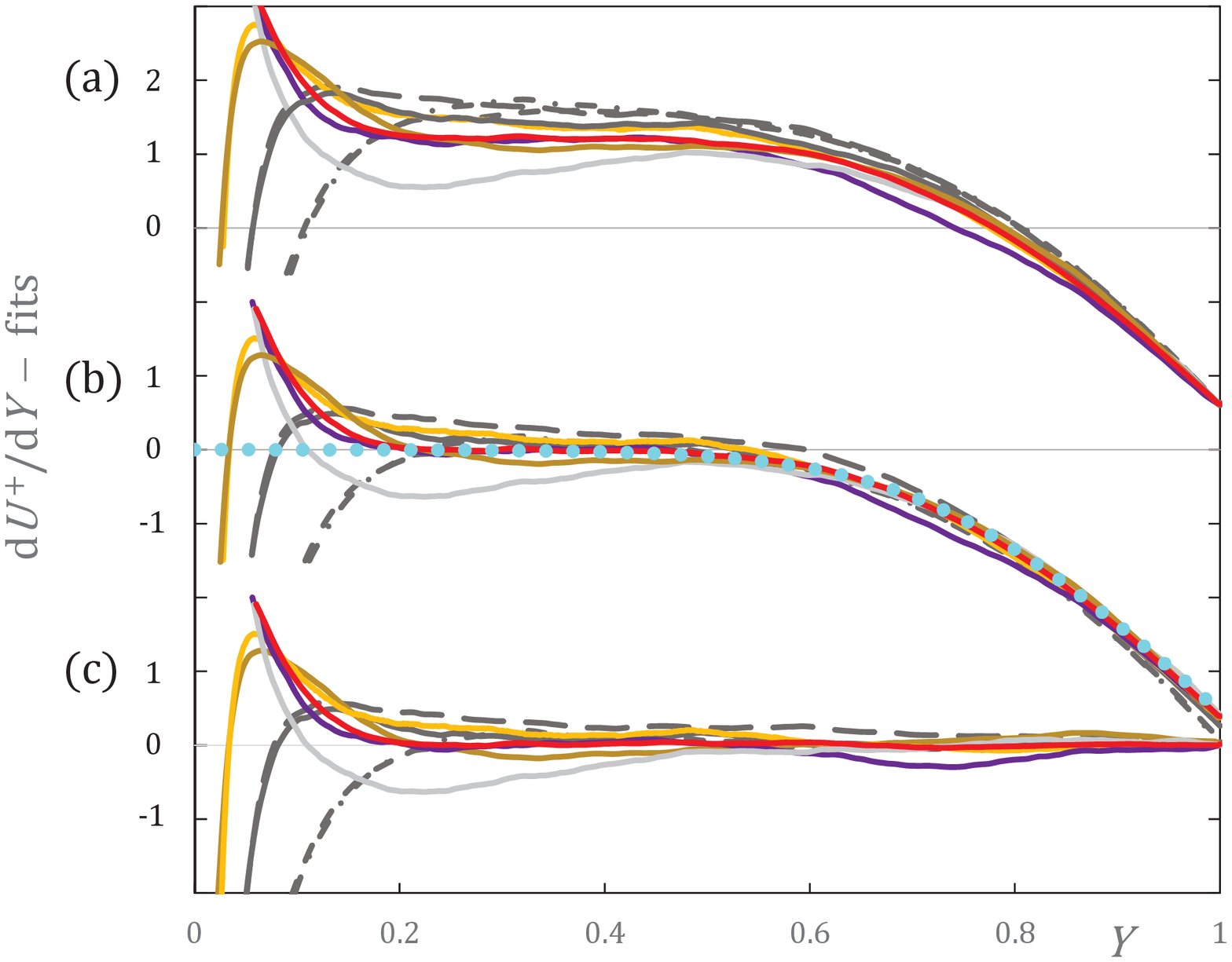}
\caption{\label{refFig2} Successive approximations of mean velocity derivative $\dd U^+_{\mathrm{DNS}}/\dd Y$ for eight channel DNS: $- \cdot -$ (grey), $\Reytau = 934$ \citep{delalamo:04}; $- - -$ (grey), $\Reytau = 1001$ \citep{LM15}; --- (grey), $\Reytau = 1995$ \citep{LM15}; --- (dark orange), $\Reytau = 3986$ \citep{Yamamoto2018}; --- (yellow), $\Reytau = 4079$ \citep{bernardini_etal_2014}; --- (red), $\Reytau = 5186$ \citep{LM15}; --- (violet),
$\Reytau = 8000$ \citep{Yamamoto2018};  --- (light grey), $\Reytau = 10049$ \citep{HoyasOberlack2022}.
Panel (a): $\dd U^+_{\mathrm{DNS}}/\dd Y - (0.417\,Y)^{-1}$; Panel (b): profiles in panel (a) minus constant $(1.15 + 380\,\Reytau^{-1})$ , $\bullet \bullet \bullet$ (blue), wake fit $[\kappa^{-1} + 1.15 + 380\,\Reytau^{-1}] (\dd W/\dd Y)$ (equations \ref{UY}, \ref{UYcofs}); Panel (c): profiles in panel (b) minus wake fit.}
\end{figure}

Starting with the velocity derivative has the advantage that, for the correct $\kappa$, the term $\dd U^+_{\mathrm{DNS}}/\dd Y - (\kappa\,Y)^{-1}$ becomes locally constant in the overlap region, irrespective of the value of $S_0$ in equation (\ref{UY}). This is clearly seen in figure
\ref{refFig2}(a), where $(\dd U^+_{\mathrm{DNS}}/\dd Y) - (\kappa\,Y)^{-1}$ is constant in the range $0.2 \lesssim Y \lesssim 0.45$ for $\kappa = 0.417$ and Reynolds numbers beyond around 2,000.

The choice of $\kappa = 0.417$ fits the profile of \citet{LM15} for $\Reytau = 5186$, considered among the most reliable, particularly well and is within the estimated range of uncertainty for the $\kappa$'s deduced from centerline velocities in \citet[][fig. 8]{Monk17} and \citet[][fig. 6]{Monk21}.
The only profile in figure \ref{refFig2}, which is not well fitted by $\kappa = 0.417$, is the $\Reytau = 10,049$ profile of \citet{HoyasOberlack2022}. Possible reasons for this discrepancy are discussed in appendix \ref{appB}.

In a second step from parts (a) to (b) of figure \ref{refFig2}, the constants in equation ({\ref{UY}), $[1.15 + 380\,\Reytau^{-1}]$, are subtracted, showing that the $\mathcal{O}(\Reytau^{-1})$ correction consistently reduces the spread between profiles of different $\Reytau$. In the last step from figure \ref{refFig2}(b) to \ref{refFig2}(c), the derivative $[\kappa^{-1} + 1.15 + 380\,\Reytau^{-1}] (\dd W/\dd Y)$ of the wake profile is subtracted, demonstrating the quality of the outer fit (\ref{UY}, \ref{UYcofs}).

How much one can be led astray when deducing K\'arm\'an ``constants'' from an inappropriate region of the indicator function (\ref{Xi}) is demonstrated with figure \ref{refFig3}, which compares the outer fit of $\Xi$, obtained from equations (\ref{UY}, \ref{UYcofs}), to the $\Xi_{\mathrm{DNS}}$ of the six highest $\Reytau$ cases of figure \ref{refFig2}.

\begin{figure}
\center
\includegraphics[width=0.65\textwidth]{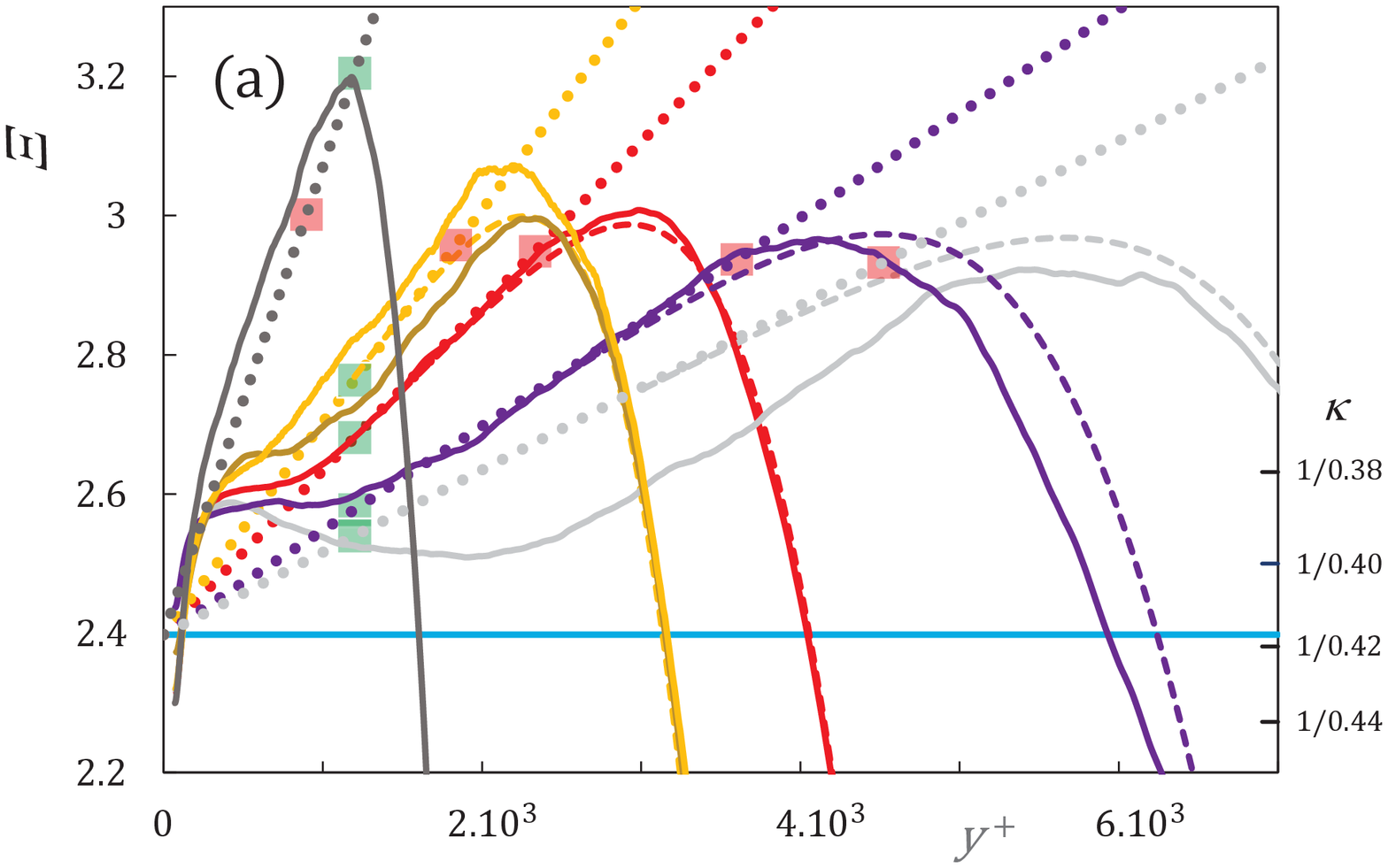}
\includegraphics[width=0.65\textwidth]{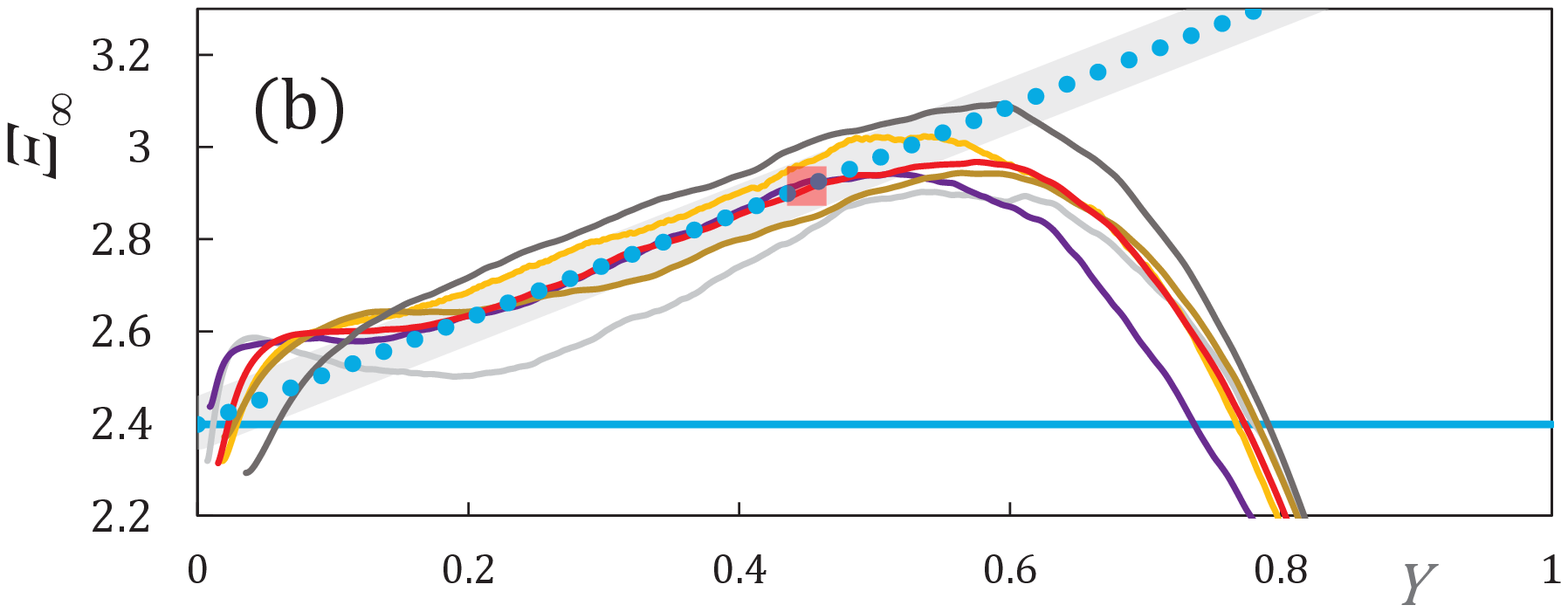}
\caption{\label{refFig3} Channel indicator functions $\Xi$ from DNS versus $y^+$ and $Y$, for the six highest $\Reytau$ of figure \ref{refFig2} ($\Reytau = 1995$ and up). Same color scheme as in figure \ref{refFig2}.
Panel (a): ---, total $\Xi$ from DNS; $\bullet \bullet \bullet$, linear overlap fits $[\kappa^{-1} + (S_0 + S_1\,\Reytau^{-1})\,Y]$ ;
 - - -, complete outer fits of $\Xi$ for $\Reytau = 4079$ and up (equations \ref{UY}, \ref{UYcofs}). --- (blue), $(1/0.417)$. $\blacksquare$ (green), $y^+ = 1200$ marking the approximate start of the overlap; $\blacksquare$ (red), $Y = 0.45$ marking the approximate end of the overlap. Note that for the lowest $\Reytau = 1995$, the overlap ends before it starts! \newline
Panel (b): Indicator functions in panel (a) corrected for finite $\Reytau$ effects according to equation (\ref{UY}, \ref{UYcofs}), i.e. $\Xi_\infty = \Xi - S_1\,Y\,[1- \dd W/\dd Y]/\Reytau$ ;
$\bullet \bullet \bullet$ (blue), leading order linear fit $(1/0.417) + 1.15\,Y$; grey band, variation of linear fit for $0.407 \leq \kappa \leq 0.427$}
\end{figure}

As already stated in the introduction, the outer expansion of $\Xi$ contains the complete (2$\mathcal{O}$inner/1$\mathcal{O}$outer) overlap (\ref{XiOL}), consisting of log law plus the linear term. Therefore, the near-wall deviations of the $\Xi_{\mathrm{DNS}}$ from their linear outer fits, i.e. the ``humps'' or ``bulges'' on top of the linear fits in figure \ref{refFig3}(a), seen for $y^+ \lessapprox 10^3$ necessarily belong to the inner expansion and not to the overlap. In particular the short, near-horizontal portions of $\Xi$ within these ``humps'', seen around $y^+$ of 500 in figure \ref{refFig3}(a), are not related to the overlap log law, but correspond to limited, approximately logarithmic \textit{inner} regions \citep[see also][fig. 12]{Monk21}.

The widespread association in the literature of these inner, nearly horizontal portions of $\Xi$ with the inner-outer overlap has fueled years of controversy about the differences between $\kappa$'s determined from these features and from the $\Reytau$-dependence of the centerline velocity, discussed in the introductory section \ref{sec1}. In addition, it has led many authors to place the overlap layer in channels and pipes too close to the wall. To just cite a carefully documented example, \citet[][table 2 and fig. 3]{LM15} estimated, for $\Reytau = 5186$, a $\kappa$ between 0.384 and 0.387 from the near-wall ``hump'' of $\Xi$, tantalizingly close to the well established $\kappa$ of 0.384 for ZPG TBLs, reported by \citet{MCN07} and \citet{NagibChauhan2008}, but significantly different from the centerline $\kappa$'s for the same data, shown in figure 8 of \citet{Monk17}, for instance.

Figure \ref{refFig3}(a) also shows the boundaries of the (2$\mathcal{O}$inner/1$\mathcal{O}$outer) overlap, defined as the locations, where the difference between $\Xi_{\mathrm{DNS}}$ and the linear overlap fit $[\kappa^{-1} + (S_0 + S_1\,\Reytau^{-1})\,Y]$ - the dotted lines in figure \ref{refFig3}(a) - falls below a set value, taken here as 0.02 . This choice results in an overlap starting at $y^+ \approxeq 1200$ and ending at $Y \approxeq 0.45$, shown by green and red squares in figure \ref{refFig3}.
Note, that with the above criterion, the overlap for $\Reytau = 1995$ ends before it starts, which means that inner and outer expansions are not yet sufficiently separated to reveal the functional form of the overlap.

In figure \ref{refFig3}(b), the $\mathcal{O}(\Reytau^{-1})$ contributions to the $\Xi$ of figure \ref{refFig3}(a), fitted by $S_1\,Y\,[1- \dd W/\dd Y]/\Reytau$ (see equations \ref{UY}, \ref{UYcofs}), have been subtracted to approximate the infinite Reynolds number limit $\Xi_\infty$ of $\Xi$.
What is somewhat surprising in this figure \ref{refFig3}(b) are the remaining rather large differences between the $\Xi_\infty$'s. The usual explanation is that, in terms of $y^+$, $\Xi$ is the product of a small and a large number. However, this is not so in terms of $Y$, which means that in current DNS practice, the outer part of the flow receives less, and possibly not enough attention compared to the near-wall part. Besides the size of the ``computational box'' and the numerical scheme, the computational grid is a likely prominent culprit. This hypothesis is examined in appendix \ref{appB}.

Confirmation for the above analysis of the channel overlap, as reflected by the $\Xi$'s obtained from DNS, is sought from experiments. Recognizing that experimental channels have a finite aspect ratio and a flow development region, one has to assume or hope that they provide MVPs that are reasonably close to those from DNS. Incidentally, channel DNS also use span-wise periodic boxes to approximate the infinite aspect ratio, with their width typically a small multiples of $\pi$ times the channel half-height, and the effect of ``quantizing'' the average aspect ratio of stream-wise rolls has, to the present authors knowledge, not yet been fully explored.

\begin{figure}
\center
\includegraphics[width=0.65\textwidth]{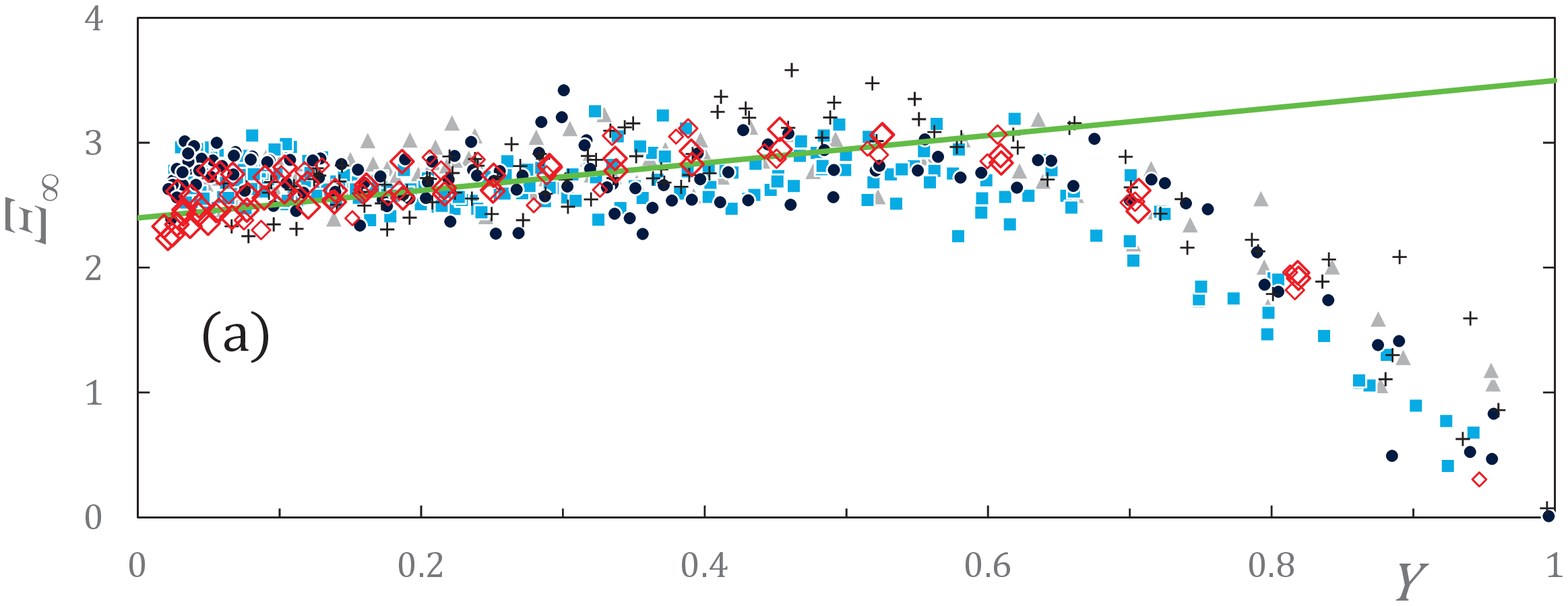}
\includegraphics[width=0.65\textwidth]{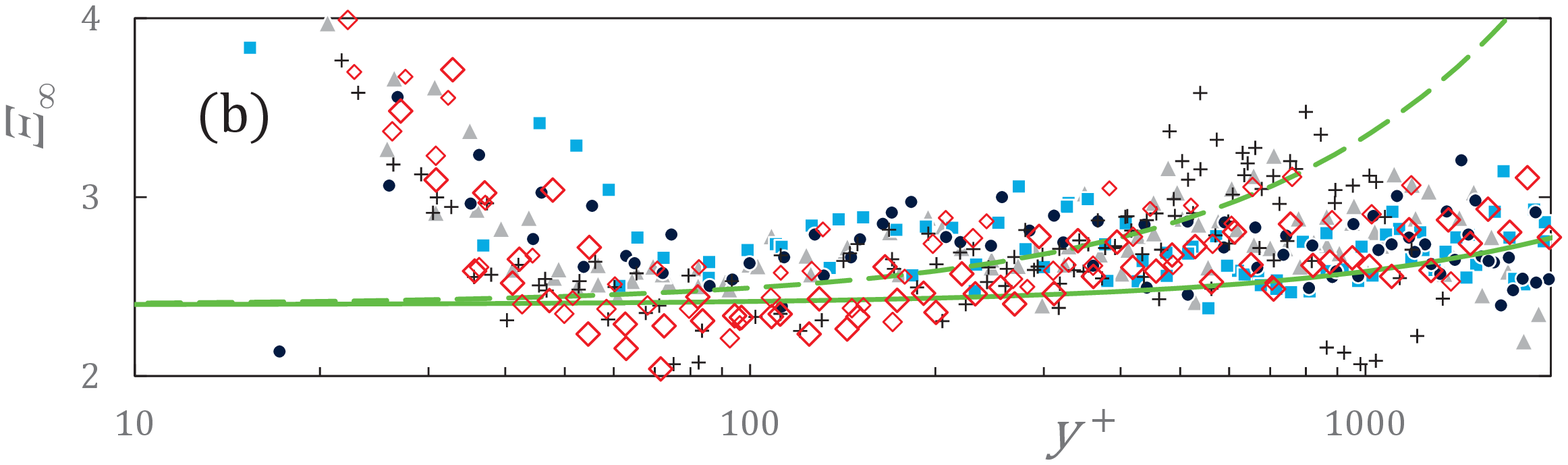}
\includegraphics[width=0.65\textwidth]{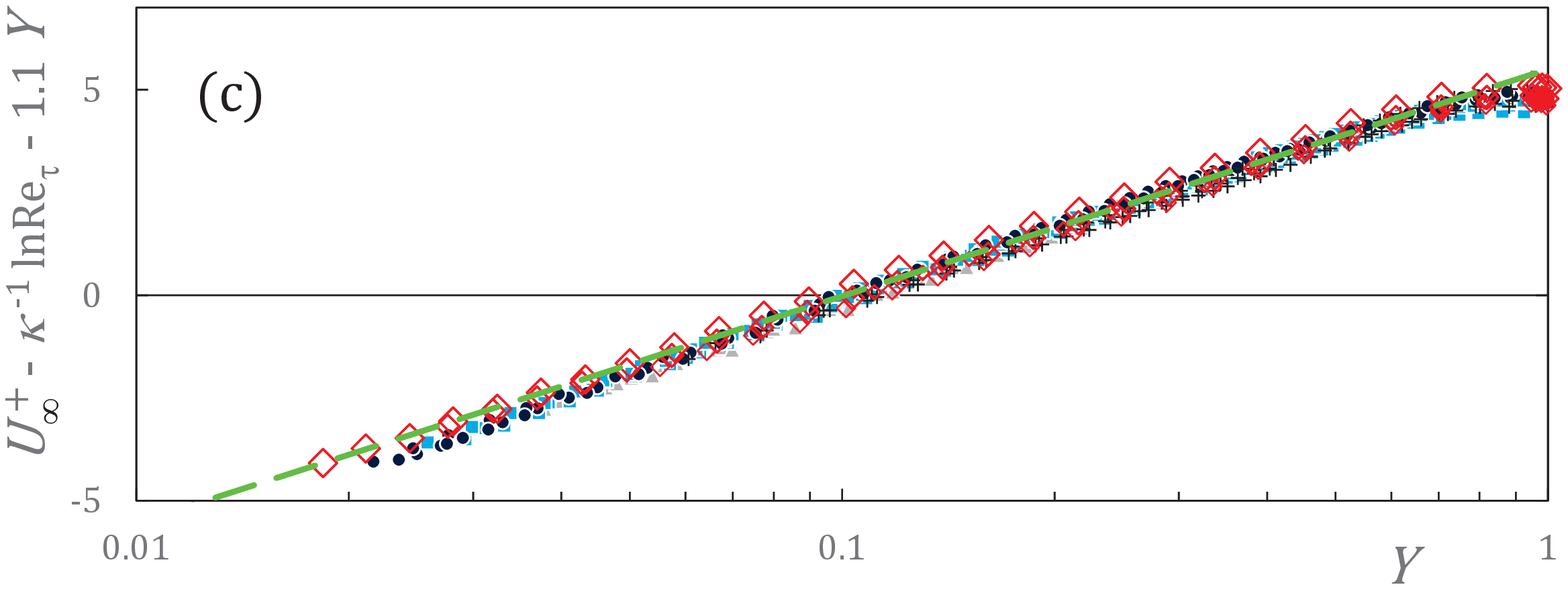}
\caption{\label{refFig4} Indicator function $\Xi_\infty(Y)$  and $U^+_\infty(Y)$
minus the linear part of the overlap
for two channel/duct experiments. The subscript ``$\infty$'' indicates that all data are corrected for finite Reynolds number effects with the $\Reytau^{-1}$ corrections in equations (\ref{UY} - \ref{Uoutcofs}).
Hot wire data of \citet{ZDN03}: + (black), $\Reytau = 1167, 1543, 1851$; $\blacktriangle$ (grey), $\Reytau = 2155,2573, 2888$; $\blacksquare$ (blue), $\Reytau = 3046, 3386, 3698, 3903$; $\bullet$ (dark blue), $\Reytau = 4040, 4605, 4783$. LDA data of \citet{SchultzFlack2013}: $\lozenge$ (red, increasing size), $\Reytau = 1010, 1956, 4048, 5895$. \newline
(a) $\Xi_\infty(Y)$: --- (green), linear fit $(1/0.417) + 1.1\,Y$ [note that the fitted $S_0 = 1.1$ is slightly reduced relative to the best DNS fit in equation (\ref{UYcofs})].
(b) Blowup of $\Xi_\infty$ versus $y^+$, with linear fits $[(1/0.417) + 1.1\,y^+/\Reytau]$ for $\Reytau = 1167$ and 5895.
(c) Corrected $U^+_\infty(Y)$ minus linear fit $[(1/0.417)\ln{\Reytau} + 1.1\,Y]$; - - - (green), resulting log law $[(1/0.417)\,\ln{Y} + 5.5]$.}
\end{figure}

The data sets for this comparison are the experimental data of
\citet{ZDN03}, who used hotwires combined with the oil film technique to determine the wall skin friction, and of \citet{SchultzFlack2013}, who used laser Doppler Anemometry (LDA). These data were obtained in channels of aspect ratio around 8, comparable to the computational box aspect ratios of the DNS in figure \ref{refFig3}. Their analysis is shown in figure \ref{refFig4}. The top two panels (a) and (b) of figure \ref{refFig4} show $\Xi_\infty$, equal to the full $\Xi$ from experiment minus the $\Reytau^{-1}$ corrections in equations (\ref{UY}, \ref{UYcofs}) versus $Y$, and an enlarged view of the wall region versus $\log{y^+}$, respectively. $\Xi_\infty$ in panel (a) shows considerable scatter due to the differentiation of experimental MVPs, but the linear trend is obvious between $Y \approx 0.2$ and $\approx 0.5$ with a slope of $S_0 = 1.1$, slightly less than the slope of 1.15 educed from channel DNS. Panel (b) shows the near-wall behavior of $\Xi_\infty$ and clearly reveals the ``hump'' on top of the linear fit $1.1\,Y$ in the data of \citet{ZDN03}, between $y^+\approx 100$ and close to $10^3$, similar to the ``humps'' in the DNS of figure \ref{refFig3}. However, for unknown reasons, the data of \citet{SchultzFlack2013} lack such a ``hump''.

The third panel (c) of figure \ref{refFig4} is the ``lynch pin'' of the data analysis, demonstrating that, after removing the linear overlap term, a clear log-law $[(1/0.417)\,\ln{\Reytau} + 5.5]$ emerges up to $Y \approx 0.5$ . The ``hump'' below $Y \approx 10^3\,\Reytau^{-1}$, seen in figure \ref{refFig4}b for the data of \citet{ZDN03}, corresponds in figure \ref{refFig4}c to the data which start to fall below the log-law fit, i.e. onto a slope of higher $\kappa^{-1}$.

At first sight, one might think that figure \ref{refFig4}c contains no new information, since $\kappa$ is already used in the linear fit of $\Xi_\infty$. However, \textit{only the slope} of $\Xi_\infty$ is used and, when subtracting $[\kappa^{-1}\,\ln{\Reytau}]$ ~from $U^+{\mathrm{corr}}$, a wrong $\kappa$ only produces vertical $\Reytau$-dependent shifts of the data sets, without affecting their logarithmic slope. Hence, this new method to determine the best fit $\kappa$ in the presence of a linear overlap term is both robust and reliable. This assessment is supported by the uncertainty estimates in section 1 of the Supplementary Material, and will be confirmed by the analogous analysis of the Superpipe data in the next section \ref{sec3}

\section{\label{sec3}The overlap of the indicator function in pipes}

Starting again with DNS indicator functions for pipe flow, figure \ref{refFig5} has the same general shape as for the channel, with a rather clear linear region at higher $\Reytau$, but a slope of about twice the slope seen in figure \ref{refFig3} for the channel. In contrast to the channel DNS, where the leading order overlap of $\Xi_\infty$ is quite well fitted by the leading order fit $(1/0.417) + 1.15\,Y$ for of all but one profile of figure \ref{refFig3}(b), the $\Xi$'s for the pipe in figure \ref{refFig5} show more substantial differences between the DNS. One likely reason is the difference of computational schemes - finite differences for the profiles of \citet{pirozzoli_pipe2021} and spectral elements for those of \citet{ElKhoury2013} and \citet{Yao2023}. In addition, the handling of the centerline grid singularity and the order of the numerical scheme may also have contributed to these differences.

As seen in figure \ref{refFig5}, the linear portion of the data of \citet{pirozzoli_pipe2021} for $\Reytau = 6019$ is quite well fitted by $[(1/0.385) + 1.95\,Y]$, while the most recent data of \citet{Yao2023} are better fitted by $[(1/0.425) + 2.75 Y]$ in the range $0.2 \leq Y \leq 0.5$ (see also figure \ref{refFig12} in appendix \ref{appB}). These discrepancies between pipe DNS are clearly more serious than in channels and do not allow us to determine a pipe $\kappa$ from the DNS results with any confidence.

As a consequence, no attempt has been made to determine finite Reynolds number corrections for pipe MVPs, analogous to those in equations (\ref{UY} - \ref{Uoutcofs}) for the channel.
Hence, the functional form of the pipe overlap and centerline velocities in terms of the outer variable $Y$ simplify to
\begin{eqnarray}
&U^+\arrowvert_{\mathrm{pipe\,OL}} = \frac{\ln{Y}}{\kappa} + \frac{\ln{\Reytau}}{\kappa} + B_0 + S_0\,Y  \label{UOLpipe}\\
&U^+\arrowvert_{\mathrm{pipe\,CL}} = \frac{\ln{\Reytau}}{\kappa} + B_0 + S_0 + W(Y=1)\quad ,  \label{UCLpipe}
\end{eqnarray}
where $W(Y)$ is the pipe wake function.

\begin{figure}
\center
\includegraphics[width=0.60\textwidth]{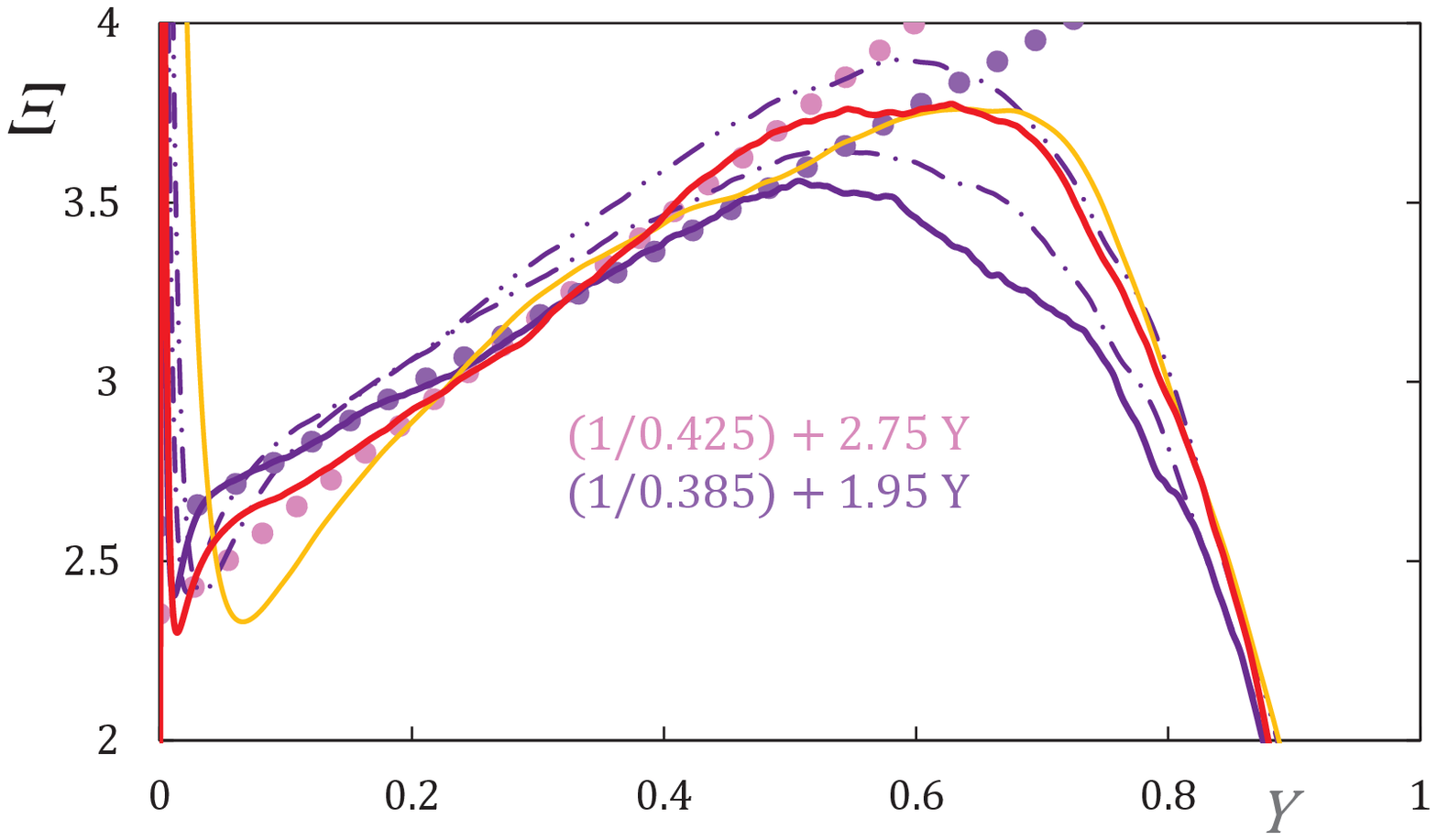}
\caption{\label{refFig5} Indicator functions $\Xi$ for selected pipe DNS (not corrected for finite $\Reytau$): --- (yellow), $\Reytau = 999$ of \citet{ElKhoury2013}; $-\cdot\cdot -$, $-\cdot -$, --- (violet), $\Reytau =$ 1976, 3028 and 6019 of \citet{pirozzoli_pipe2021}; --- (red), $\Reytau = 5197$ of \citet{Yao2023}. $\bullet \bullet \bullet$ (violet), linear fit $(1/0.385) + 1.95\,Y$ of $\Reytau = 6019$ profile; $\bullet \bullet \bullet$ (pink), linear fit $(1/0.425) + 2.75\,Y$ of $\Reytau = 5197$ profile. }
\end{figure}

Turning to experiments, we focus on the Superpipe data \citep{ZS98,McK_thesis,Pitot13} that are probably the most scrutinized experimental data in turbulence history. Starting with the centerline, the Superpipe data are the only recent data set which cover nearly two decades of high Reynolds numbers ($\geq 10^4$), allowing
a rather reliable estimate of $\kappa$ from the $\Reytau$ dependence of centerline velocity. While the near-wall Pitot data have been the object of numerous challenges and corrections \citep[see for instance][]{VinNagib15}, the centerline velocities have remained virtually unaffected and allows $\kappa$ to be determined from $U^+_{\mathrm{CL}}(\Reytau)$ (equation \ref{UCLpipe}). In \citet[][fig. 4]{Monk17} the quality of the fit with the original $\kappa = 0.436$ of \citet{ZS98} was found to be comparable to the one with $\kappa = 0.42$, deduced by \citet{McK_thesis}, but the comparative study of \citet{NagibTSFP10} suggests, that the pipe centerline $\kappa$ is closer to the original $\kappa = 0.436$ of \citet{ZS98} than to 0.42.

\begin{figure}
\center
\includegraphics[width=0.65\textwidth]{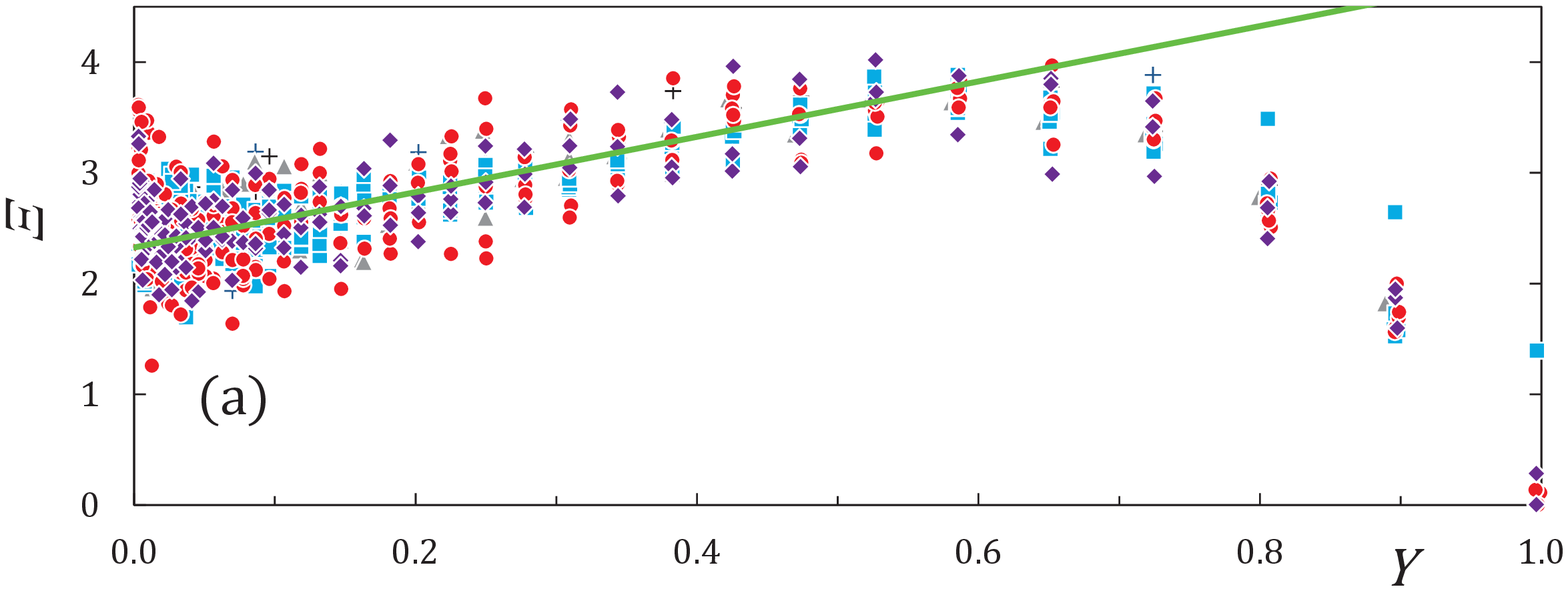}
\includegraphics[width=0.65\textwidth]{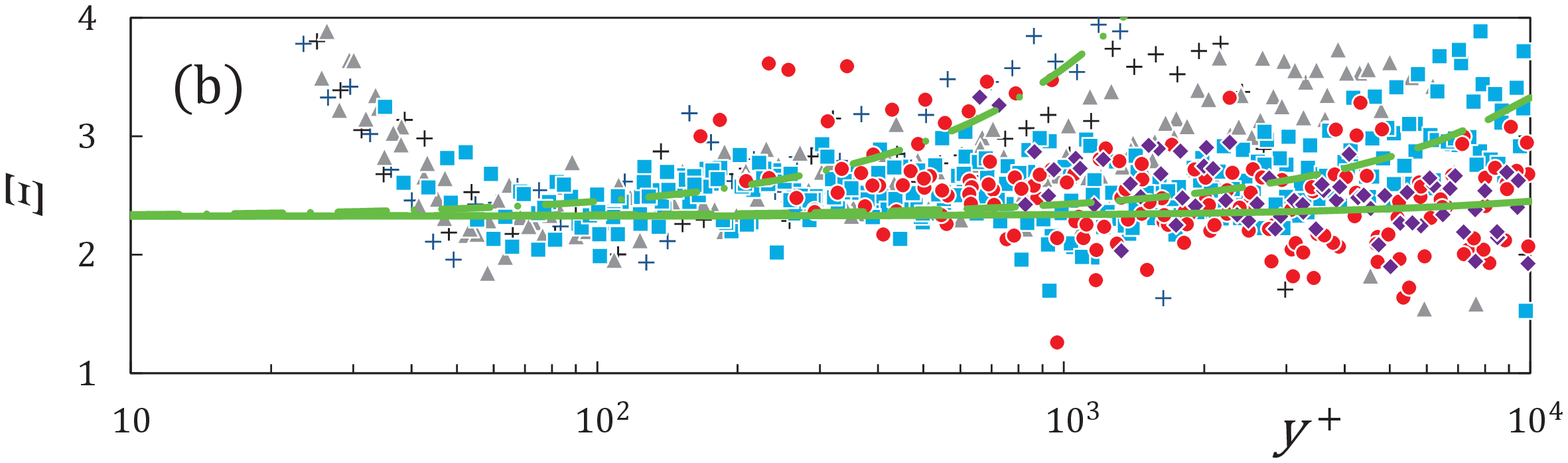}
\includegraphics[width=0.65\textwidth]{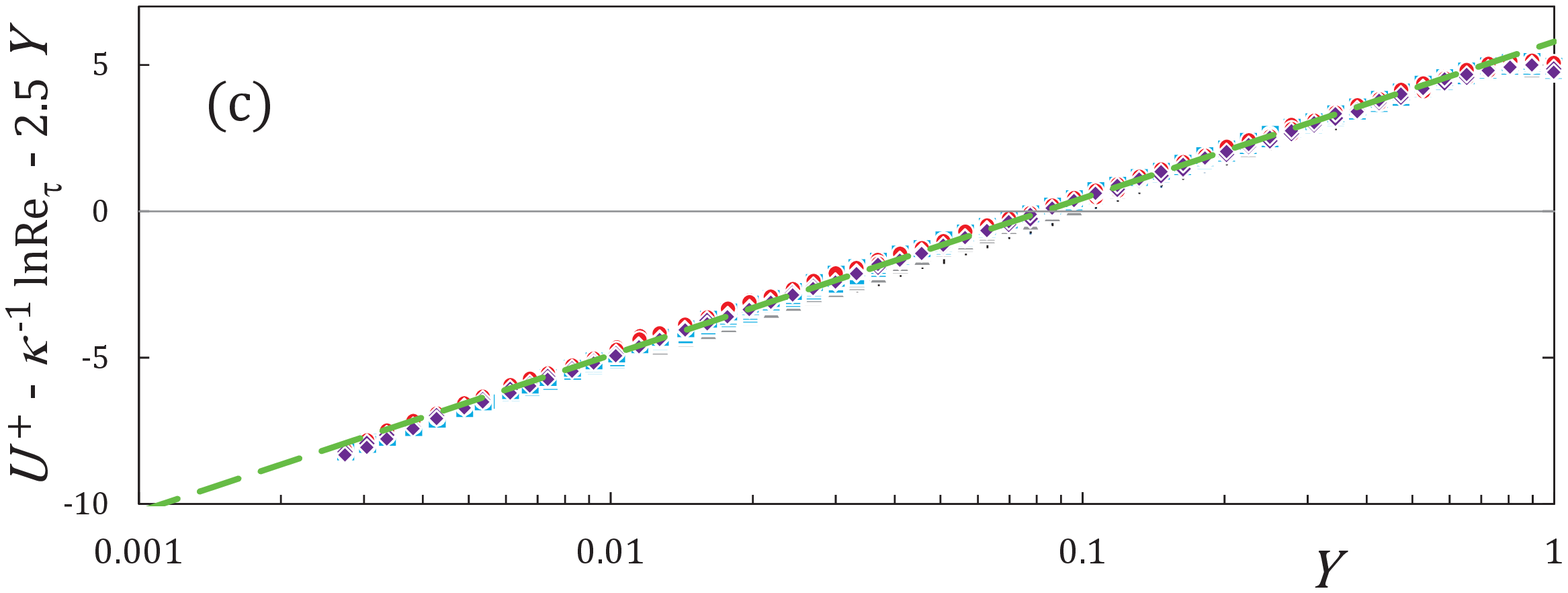}
\caption{\label{refFig6} Indicator function $\Xi(Y)$  and $U^+(Y)$ minus linear part of overlap (equation \ref{UOLpipe})
for the Superpipe data of \citet{McK_thesis,Pitot13}.  + (black), $\Reytau < 5.10^3$; $\blacktriangle$ (grey), $5.10^3 < \Reytau < 10^4$; $\blacksquare$ (blue), $10^4 < \Reytau < 5.10^4$; $\bullet$ (red), $5.10^4 < \Reytau < 2.10^5$; $\blacklozenge$ (purple), $2.10^5 < \Reytau < 5.3\,10^5$.\newline
(a) $\Xi(Y)$: --- (green), linear fit $(1/0.433) + 2.5\,Y$.
(b) Blowup of $\Xi$ versus $y^+$, with linear fits $[(1/0.433) + 2.5\,y^+/\Reytau]$ for $\Reytau = 2000$, 25,000 and 250,000.
(c) $U^+(Y) - [(1/0.433)\ln{\Reytau} + 2.5\,Y]$; - - - (green), resulting log law $(1/0.433)\,\ln{Y} + 5.8$.}
\end{figure}

Since pipe flow is pressure driven like channel flow, it is natural to use the methodology of section \ref{sec2} to determine the pipe overlap parameters $\kappa$ and $S_0$ in equation (\ref{UOLpipe}).
The Superpipe data, corrected according to \citet{Pitot13}, are shown in figure \ref{refFig6} in the same format as the channel data in figure \ref{refFig4}, but \textit{without subtracting finite Reynolds number corrections} from the data. The indicator function $\Xi$, shown in panel (a) of the figure, has a clear linear part, well fitted by $[(1/0.433) + 2.5\,Y]$, which extends to $Y \approxeq 0.45$. The enlarged view of $\Xi$ in panel (b) versus $y^+$ shows again the ``hump'' of $\Xi$ between $y^+ \approx 10^2$ and about $10^3$, similar to the ``hump'' in one set of experimental channel profiles and in all channel and pipe $\Xi$'s from DNS.

The scatter of $\Xi$ in figure \ref{refFig6}a is again relatively large, due to the differentiation of experimental data, and the uncertainty of the intercept is estimated at $[0.433 \pm 0.03]^{-1}$. However, as already made clear in section \ref{sec2}, only the slope $S_0$ of $\Xi$ is needed to determine $\kappa$ from the logarithmic slope in figure \ref{refFig6}c. This last figure shows a remarkable data collapse onto the log law $[(1/0.433)\,\ln{Y} + 5.9]$ over about half the pipe radius, with an estimated uncertainty in $\kappa$ of $\pm \,0.01$. Furthermore, no Reynolds number trend of the linear slope $S_0$ can be detected over the entire range of the Superpipe Reynolds numbers ! A more detailed uncertainty analysis can be found in section 2 of the Supplementary material.

The alternative approach of first determining $\kappa$ from the centerline velocity (\ref{UCLpipe}) has recently been made possible by an upgrade of the CICLoPE pipe \citep[see][for a description of the facility]{NagibTSFP10}, in which reliable hotwire MVPs are available. Knowing $\kappa$, $L_0$ and $B_0$ can thus be obtained by a linear fit to the overlap MVP (\ref{UOLpipe}) minus the log-law.

Significantly, the slope of the linear overlap term, obtained from the Superpipe profiles of this section, is roughly twice the slope in the channel overlap profile. This supports the basic finding of \citet{Luchini17}, that, for sufficiently small pressure gradients, $S_0$ is proportional to the pressure gradient parameter $\beta \equiv -(\widehat{\mathcal{L}}/\widehat{\tau_w}) (\dd \widehat{p}/\dd \widehat{x})$, equal to 1 and 2 for channels and pipes, respectively. His dimensional analysis was however unnecessarily constrained, as discussed in appendix \ref{appC}.

\section{\label{sec4}Comparison to the outer MVP in the ZPG TBL}

According to \citet{Luchini17}, the ZPG TBL is the only one of the three ``canonical'' flows considered in the present paper, in which the overlap is a pure log law without a linear component. However, as opposed to channel and pipe flow, the ZPG TBL is slightly non-parallel. The result is, as argued by \citet{Spalart}, that the mean advection term behaves like a non-zero pressure gradient term. It is not clear how this affects the overlap, but if a linear overlap term should result, it is too small to be seen in the top part of figure \ref{refFig7}, which shows $\Xi$ obtained from the three experimental data sets of \citet{samie_etal_2018}, \citet{jens:phd} and \citet{NCM07} (note that for the latter two data sets, the $\Reytau$ values have been rescaled to match the definition of boundary layer thickness by \citeauthor{samie_etal_2018}).

The striking difference to figures \ref{refFig4}a and \ref{refFig6}a is the clean (within experimental scatter) overlap log law, with the widely accepted best fit $\kappa = 0.384$ of \citet{MCN07}, which ends abruptly at the outer wall distance of $Y = 0.11\,$.

\begin{figure}
\center
\includegraphics[width=0.7\textwidth]{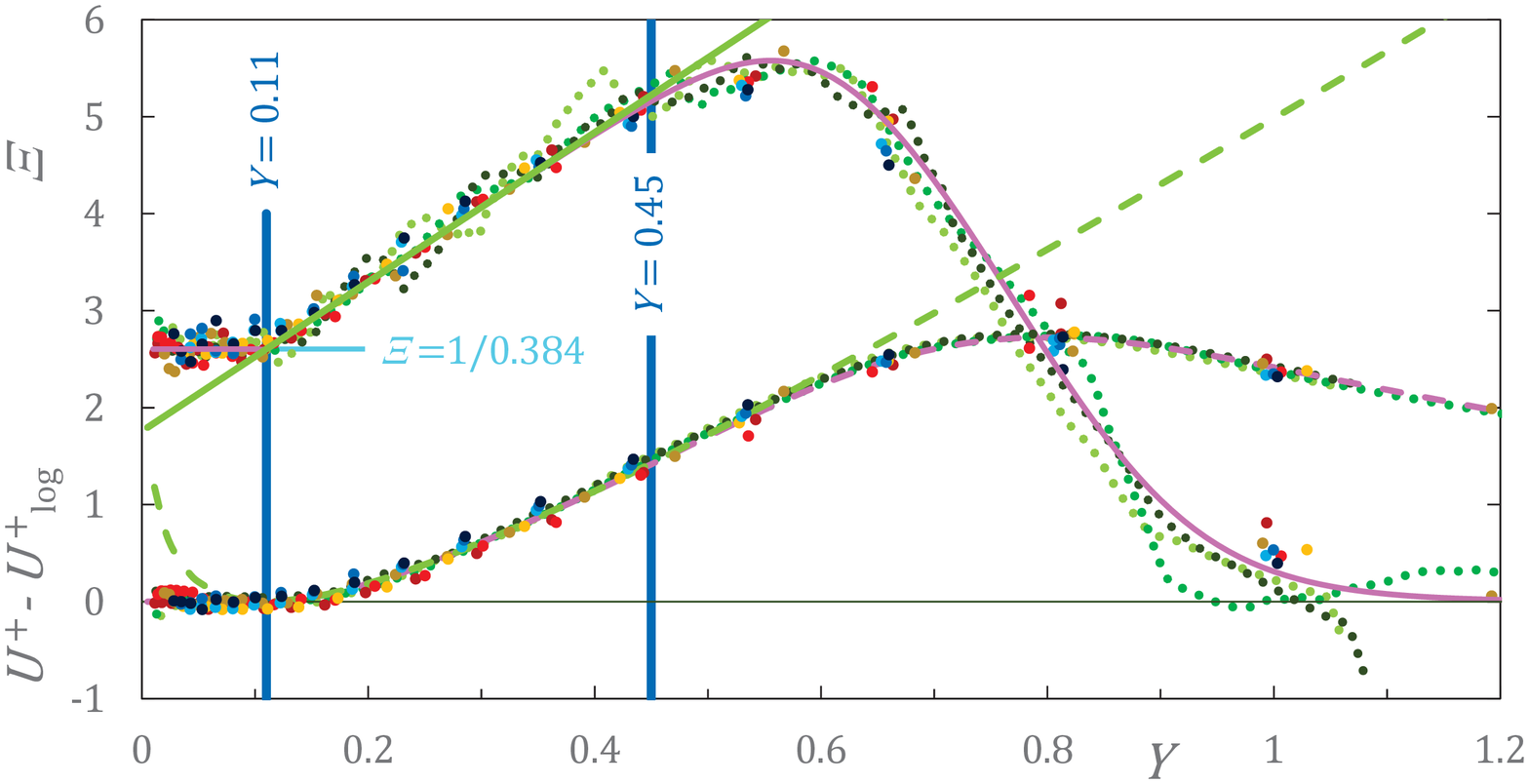}
\caption{\label{refFig7} ZPG TBL
Indicator function $\Xi(Y)$ (\textit{top}) and $U^+(Y)$ minus log law $[(1/0.384)\ln{y^+} + 4.17]$ (\textit{bottom}). $\bullet$ (yellow, dark yellow, red, dark red), data of \citet{samie_etal_2018} for $\Reytau = 6, 10, 14.5\, \&\, 20 \times 10^3$. $\bullet$ (light blue, blue, dark blue), data of \citet{jens:phd} for $\Reytau = 5.5, 6.6 \,\&\, 7.9 \times 10^3$. $\bullet\bullet\bullet$ (increasingly dark green), data of \citet{NCM07} for $\Reytau = 12.6, 16 \,\&\, 22.5 \times 10^3$ (for this last set, the log law constant has been increased from 4.17 to 4.32).
Fits: --- (light blue), $\Xi = 1/0.384$; --- (light green), linear part $\Xi = (1/0.384) + 7.7\,(Y-0.11)$ for $0.11 \leqq Y \lessapprox 0.45$; --- (lavender), full fit of $\Xi$ (equations \ref{Xifit},\ref{Xifitcofs}); - - - (light green), fit $7.7\,[Y - 0.11 - 0.11\,\ln{(Y/0.11)}]$ corresponding to the linear part of $\Xi$. - - - (lavender), full fit of $U^+$ - log law (num. integration of equations \ref{Xifit},\ref{Xifitcofs}).}
\end{figure}

The next part of $\Xi(Y)$ in figure \ref{refFig7}, between $Y = 0.11\,$ and $\approxeq 0.45$ is linear, with a large slope of $7.7\,$. As this linear part starts at a fixed \textit{outer} location, it has nothing to do with the overlap and its physical origin is different. One likely candidate is the entrainment of free stream fluid into the boundary layer, as discussed by \citet{Chauhan14a,Chauhan14b}, for instance. The part of $\Xi$ beyond $Y \approxeq 0.45$ represents the transition to the free stream where $\Xi = 0\,$.

Both logarithmic overlap and linear part of $\Xi$ are seen in the bottom part of figure \ref{refFig7} to provide, upon integration, an excellent outer fit of the MVP up to $Y \approxeq 0.45\,$. As the clear division of the outer $\Xi$ into constant and linear parts appears more physical than the classical wake formulation of \citet{Coles56}, the following full outer fit is proposed:
\begin{eqnarray}
& \Xi(Y)\arrowvert_{\mathrm{outer\,fit}} = \left\{\frac{1}{\kappa} + \Lambda\,(Y - Y_{\mathrm{break}})\,\mathcal{H}(Y-Y_{\mathrm{break}})\right\}\left\{1 - Y\,\frac{\dd W}{\dd Y}\right\} \label{Xifit} \\
& \mathrm{with}\,\, \mathcal{H}\,\, \mathrm{the \,\,Heaviside \,\,function},  \,
\kappa = 0.384, \, \Lambda = 7.7, \, Y_{\mathrm{break}} = 0.11 \quad \mathrm{and}  \nonumber \\
& Y\,\frac{\dd W}{\dd Y} = \left\{\frac{1}{19}\,\ln{\left[1 + e^{19(Y- 0.59)}\right]} - \frac{1}{15}\,\ln{\left[1 + 2\,e^{15(Y- 0.92)}\right]}\right\} \times \nonumber \\
& \quad \times\,\left\{0.92 - 0.59 - \frac{\ln{2}}{15}\right\}^{-1}
\label{Xifitcofs}
\end{eqnarray}
where $Y$ is defined as in \citet{samie_etal_2018}. Equation (\ref{Xifit}, \ref{Xifitcofs}), upon numerical integration, yields an excellent outer fit of the MVP in ZPG TBLs, as demonstrated by the dashed lavender curve in the lower part of figure \ref{refFig7}.

\section{\label{sec5}Conclusions}

The main conclusion of the present study is
that the overlap of the MVP in channels, pipes and ZPG TBL's is not universal. This non-universality includes the overlap parameters, as well as the start or end locations.
This non-universality should however not come as a surprise, as the MVP overlap provides the transition between the near-universal part of the profile in the inner, near-wall region and the geometry-dependent outer part of the profile. How close to universal the inner parts of the MVP in ZPG TBLs, channels and pipes really are, still remains to be investigated more thoroughly. At any rate, they could only be strictly universal in the limit of $\Reytau \to \infty$, since the Taylor expansion of $U^+$ about the wall contains the higher order term $\beta\,(2\,\Reytau)^{-1}(y^+)^2$ which depends on the pressure gradient parameter $\beta$ \citep[see e.g.][sec. 3.3]{Monk21}.\newline

\begin{figure}
\center
\includegraphics[width=0.49\textwidth]{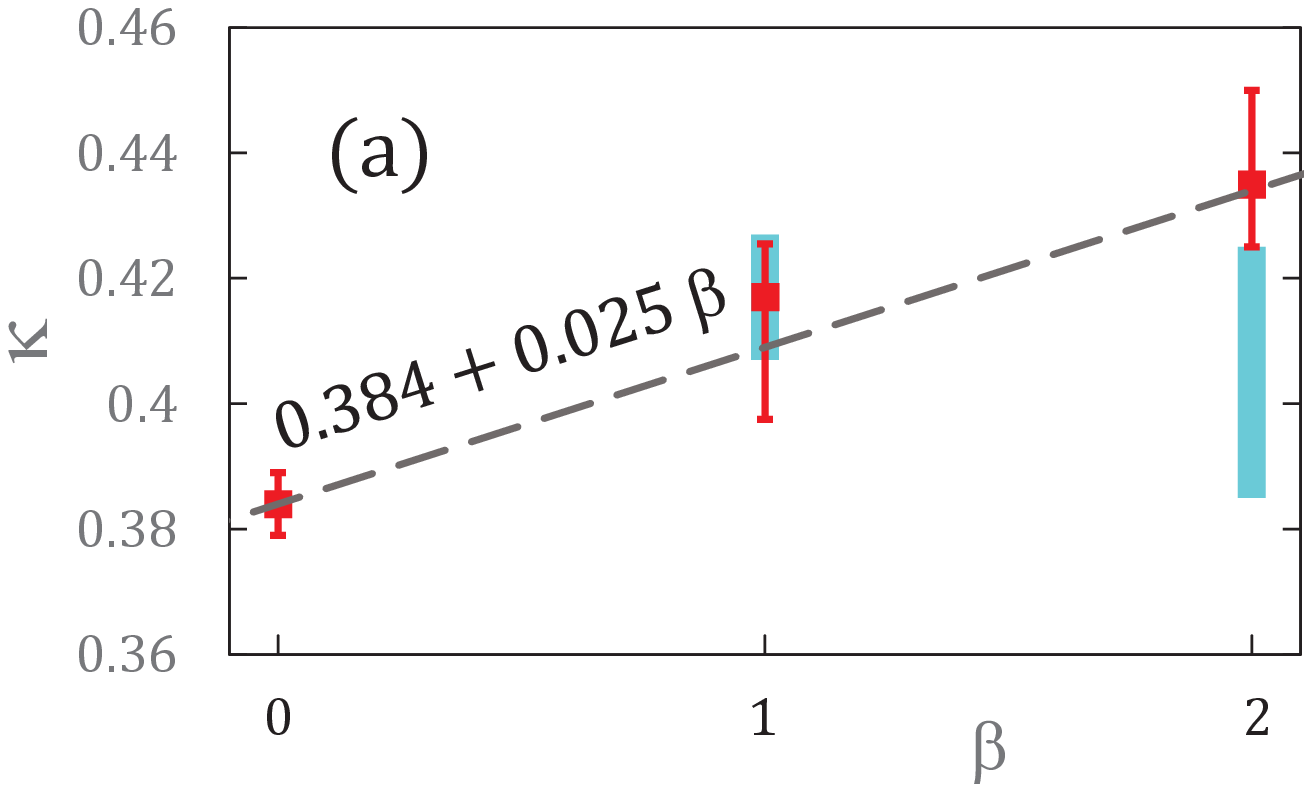}
\includegraphics[width=0.49\textwidth]{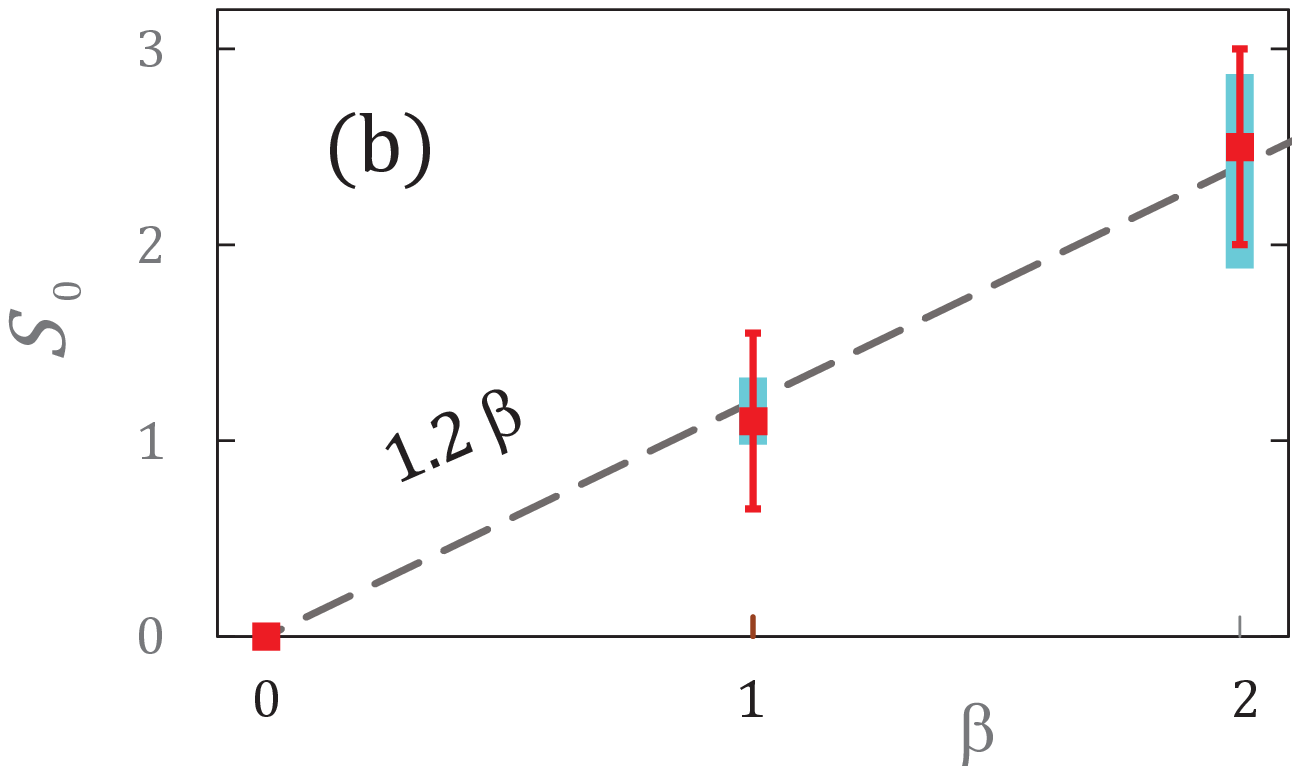}
\caption{\label{refFig8} Dependence of the overlap parameters $\kappa$ (panel a) and the slope of the linear term $S_0$ (panel b) on the pressure gradient parameter $\beta \equiv -(\widehat{\mathcal{L}}/\widehat{\tau_w}) (\dd \widehat{p}/\dd \widehat{x})$, equal to 1 and 2 for channel and pipe. Blue vertical bars, range of values deduced from the DNS of figures \ref{refFig3} and \ref{refFig5}; red $\blacksquare$, baseline fits of the experimental data of figures \ref{refFig7} (ZPG TBL), \ref{refFig4} (channel) and \ref{refFig6} (pipe) with uncertainty estimates elaborated in the Supplementary Material. --- (grey), tentative linear fits.}
\end{figure}

\noindent \textit{Specific results of the present analysis are} :
\begin{enumerate}
\item The overlap in channels and pipes does not start until $y^+ \approxeq \mathcal{O}(10^3)$, as already discussed by \citet{Monk21}. This follows from the \textit{outer} expansion of the indicator function $\Xi$ which must contain the overlap and, for small $Y$, is a simple linear function of $Y$.
\item The (1$\mathcal{O}$inner/1$\mathcal{O}$outer) overlap of the MVP, i.e. the pure log law, is not useful in channel and pipe flow, since extreme Reynolds numbers are required to reveal it over an extended interval of $y^+$.
\item To remedy this problem, which is specific to channel and pipe flow, and more generally to flows with stream-wise pressure gradient, one has to resort to the (2$\mathcal{O}$inner/1$\mathcal{O}$outer) overlap, which contains, in addition to the log law, the linear term $S_0 (y^+/\Reytau) \equiv S_0\,Y$ (see also the discussion in section \ref{sec1}). This (2$\mathcal{O}$inner/1$\mathcal{O}$outer) overlap is clearly seen in channels and pipes for $\Reytau \gtrapprox 5.10^3$ and extends from  $y^+ \approx 10^3$ to $Y \approx 0.5\,$ with its center located at the intermediate variable $(y^+ Y)^{1/2} \approx 20-25$.
\item Based on these findings, a new and robust method has been developed to simultaneously extract $\kappa$ and $S_0$ from the MVP of pressure-driven flows at currently accessible $\Reytau$’s. This new method yields $\kappa$’s, which are consistent with the $\kappa$’s deduced from the leading order Reynolds number dependence $\ln{\Reytau}/\kappa$ of centerline velocities by \citet{NagibTSFP10}, \citet{Monk17} and  \citet{Monk21}, for instance.
    As discussed in section \ref{sec3}, it is also possible to first determine $\kappa$ from the $\Reytau$ dependence of the centerline velocity (equation \ref{UCLpipe}), to subtract the log-law from the overlap velocity (\ref{UOLpipe}), and to obtain $L_0$ and $B_0$ by a linear fit to the remainder.
    The present estimates for the dependence of these parameters on the pressure gradient parameter $\beta \equiv -(\widehat{\mathcal{L}}/\widehat{\tau_w}) (\dd \widehat{p}/\dd \widehat{x})$, equal to 1 and 2 for channel and pipe, are reflected in figure \ref{refFig8}.
\item As opposed to channel and pipe flows, the outer expansions of $\Xi$ and $U^+$ in the ZPG TBL feature a clear logarithmic overlap with $\kappa = 0.384$, which ends at $Y = 0.11$ (using the definition of \citet{samie_etal_2018} for the boundary layer thickness). Beyond the overlap, a linear part of the outer $\Xi$ has been identified in the interval $Y \in [0.11, 0.45]$, followed by the transition to the free stream. A new outer fit with these features has been presented in section \ref{sec4}.
\item Regarding DNS, more higher quality channel and pipe DNS at $\Reytau$'s around $10^4$ with an increased attention to the accuracy of the outer part of the flow are required to narrow down the values of the overlap parameters in channel and pipe flows and to fully clarify their asymptotic structure.
    Increasing the accuracy of MVPs should have priority over attempts to reach new record Reynolds numbers. It may also be interesting to perform DNS of high Reynolds number flows situated somewhere between channel and pipe flow, i.e. in rectangular or elliptic ducts of different aspect ratio, similar to the simulations of \citet{VSN2018} at low $\Reytau$'s.
\end{enumerate}

\begin{acknowledgements}
We are grateful to Katepalli Sreenivasan for the valuable discussions on several points of this manuscript, to Ricardo Vinuesa for sharing his expertise in DNS, which is reflected in the discussion in appendix \ref{appB}, and to Jie Yao for providing figure 5 of \citet{Yao2023} in numerical form. \newline
\end{acknowledgements}

Declaration of Interests. The authors report no conflict of interest.
\appendix
\section{Additional information on the outer expansion of the channel MVP in section \ref{sec2}}\label{appA}

In section \ref{sec2}, the step from figure \ref{refFig2}a to \ref{refFig2}b involved the subtraction of contributions of $\mathcal{O}(\Reytau^{-1})$. In \citet{Monk21}, this higher order correction for the velocity derivative was modelled by a $\sin{(\pi Y)}$ function, reproduced here in figure \ref{refFig9}. This fit has been simplified in equation (\ref{UY}). The simplified $\mathcal{O}(\Reytau^{-1})$ fit is seen in figure \ref{refFig9} to be equivalent to the previous one, except for $Y \lesssim 0.2$ where it cannot be deduced from DNS at currently available $\Reytau$'s.

\begin{figure}
\center
\includegraphics[width=0.65\textwidth]{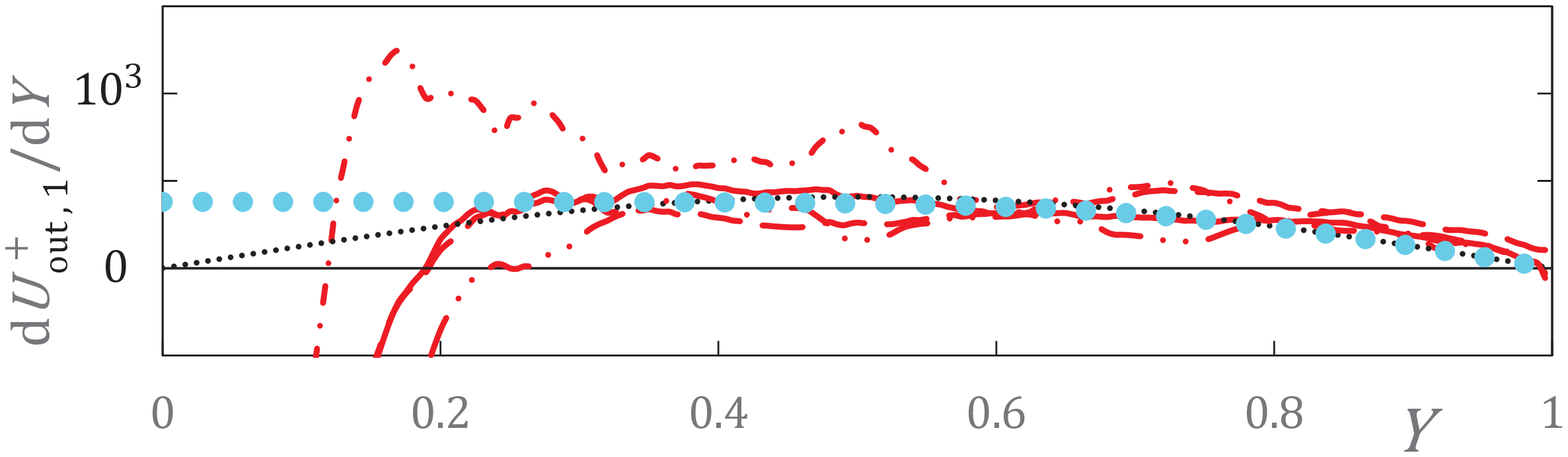}
\caption{\label{refFig9} Contributions of order $\mathcal{O}(\Reytau^{-1})$ to the channel $\dd U^+/\dd Y$, taken from figure 4b of \citet{Monk21}. $\bullet \bullet \bullet$ (blue), new fit (\ref{UY}, \ref{UYcofs}); $\cdots$ (black), previous fit in \citet{Monk21}.}
\end{figure}

\begin{figure}
\center
\includegraphics[width=0.65\textwidth]{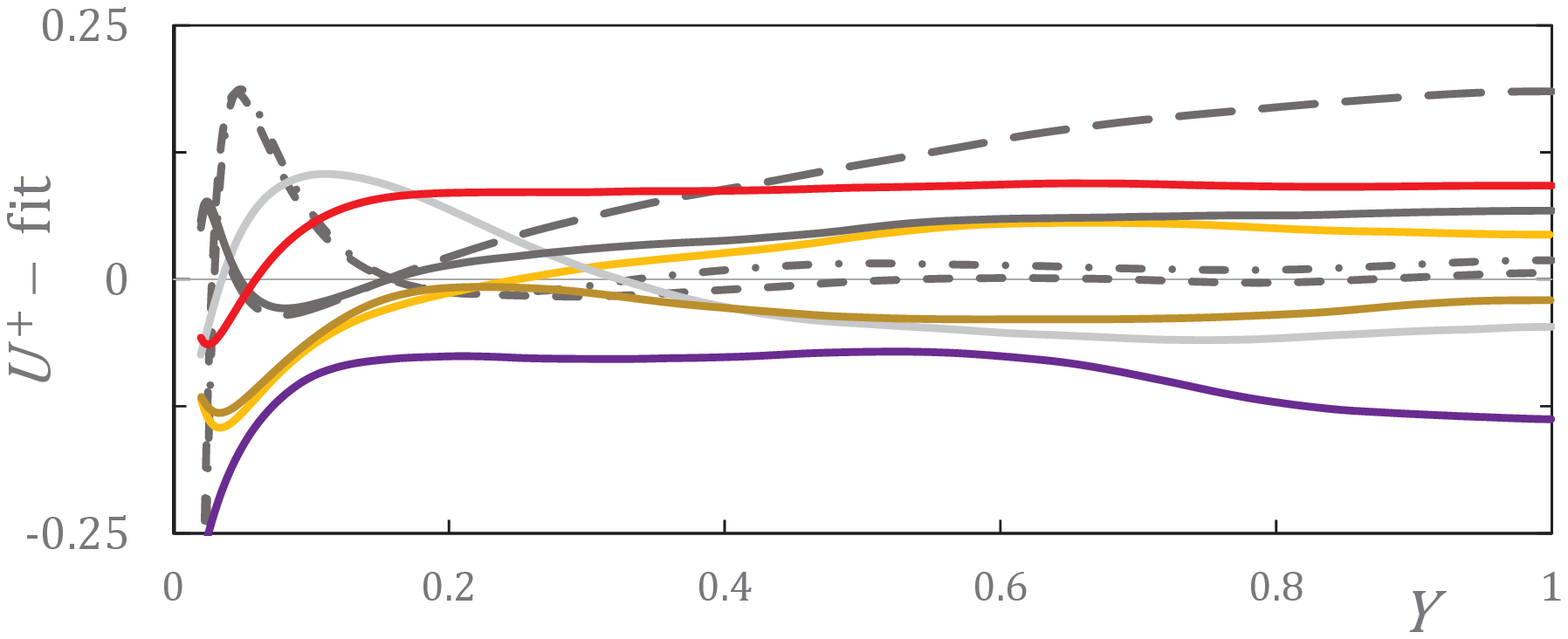}
\caption{\label{refFig10} DNS mean velocity for the channel DNS of figure \ref{refFig2}, minus the outer fit (\ref{Uout}, \ref{Uoutcofs}).
Same color scheme as in figure \ref{refFig2}.}
\end{figure}

To further validate the present outer expansion of the channel mean velocity, the
MVPs of the channel DNS used in figure \ref{refFig2} minus the outer fit (\ref{Uout}, \ref{Uoutcofs}) are shown in figure \ref{refFig10} to collapse rather well, although individual profiles could obviously be better fitted with slight adjustments of the parameters.
For $\Reytau$'s beyond $10^3$, the differences between DNS and the outer fit are below 0.5\% of centerline velocities, all the way to the inner boundary of the overlap. The latter can be estimated from figure \ref{refFig10} to be around $y^+ \approx 10^3$ (corresponding to $Y \approx 0.15$ for the Reynolds numbers in this figure), in accord with the conclusions of \citet{Monk21}. The outer limit of the overlap is seen in figure \ref{refFig2} to be located at $Y \approxeq 0.4 - 0.45$, with the exact value depending on the maximum deviation allowed between overlap and full profile.

\section{Exploring the effect of DNS grid spacing on indicator functions}\label{appB}

The differences between the indicator functions from different DNS, seen in figure \ref{refFig3}, are rather large, even after subtracting $\Reytau^{-1}$ corrections in panel (b), and call for an explanation. Here, a correlation with the grid spacing is explored in
figure \ref{refFig11}, which shows the distribution of grid spacing $\Delta y^+$ over the channel half-height for a number of channel DNS, together with two indicator functions of \citet{Yamamoto2018}.

What is striking in this figure \ref{refFig11} is the rapid increase of $\Delta y^+$, reaching 2 already at $y^+ = 100$, and reaching 10-15 on the centerline. The exceptions are the two DNS of \citet{Yamamoto2018},
where this increase is delayed to $y^+ \approx 10^3$. In the group of DNS without those of \citeauthor{Yamamoto2018}, the DNS of \citet{LM15} for $\Reytau = 5186$ has the smallest $\Delta y^+$'s, while those of \citet{HoyasOberlack2022} for $\Reytau = 10,049$ are about twice as large, which has probably contributed to the ``untypical'' $\Xi$ for this case in figure \ref{refFig3}.

Unfortunately, the two DNS of \citet{Yamamoto2018} have the largest $\Delta y^+$'s close to the wall and use a second order scheme, which possibly plays a role. Nevertheless, the distribution of $\Delta y^+$'s may contribute to the explanation for the scatter in figure \ref{refFig3}b. In figure \ref{refFig11} the $\Xi$ of these two DNS clearly change slope in the part that is linear in other DNS. The location of these slope changes, marked in the figure by large dots, correlates quite well with the location where the respective $\Delta y^+$ reach a value of around 4.

\begin{figure}
\center
\includegraphics[width=0.65\textwidth]{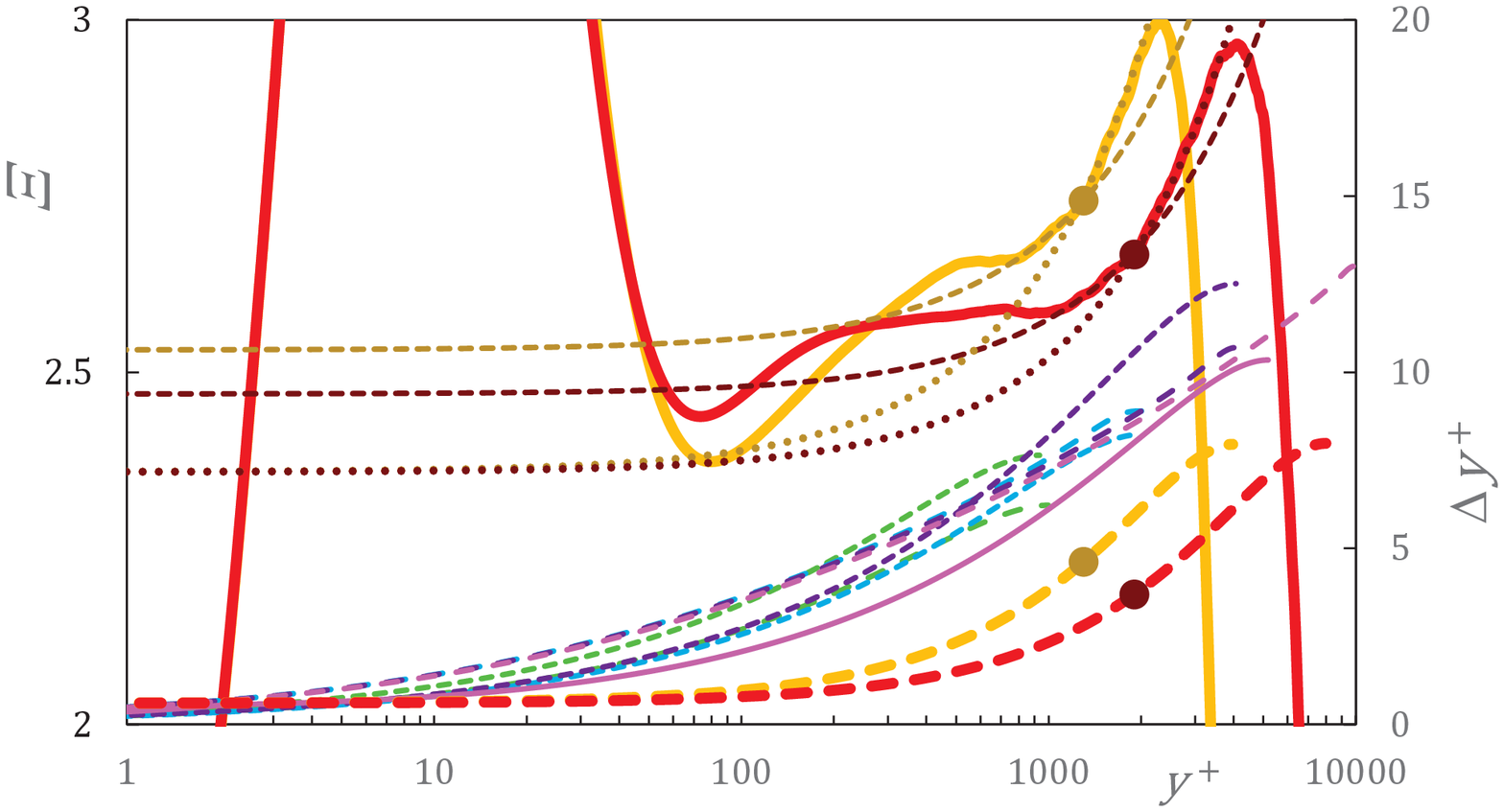}
\caption{\label{refFig11} Grid spacing $\Delta y^+$ of different channel DNS (right vertical axis) versus $y^+$ compared to two indicator functions of \citet{Yamamoto2018} (left axis).\newline
\textit{Left axis}.
--- (orange), $\Xi(\Reytau = 3986)$; $\cdot\cdot\cdot$ (light brown), linear fit $(1/0.424) + 1.2\,Y$; - - - (light brown), fit $(1/0.395) + 0.65\,Y$; $\bullet$ (light brown), switch between the two linear fits at $y^+ \approxeq 1300$. --- (red), $\Xi(\Reytau = 8000)$; $\cdot\cdot\cdot$ (dark red), linear fit $(1/0.424) + 1.3\,Y$; - - - (dark red), fit $(1/0.405) + 0.85\,Y$; $\bullet$ (dark red), switch between the two linear fits at $y^+ \approxeq 1900$.\newline
\textit{Right axis}. Grid spacing for $\Reytau = 3986$ (thick orange dashes) and 8000 (thick red dashes) of \citet{Yamamoto2018} with location of the change of linear slope of $\Xi$ indicated by bullets.
Comparison grid spacings shown for $\Reytau$ = 934 \citep[][short green dashes]{delalamo:04}, 1001 \citep[][long green dashes]{LM15}, 1995 \citep[][short blue dashes]{LM15}, 2004 \citep[][long blue dashes]{HJ06}, 4079 \citep[][long violet dashes]{bernardini_etal_2014}, 4179 \citep[][]{LJ14}, 5186 \citep[][pink solid line]{LM15}, 10046 \citep[][pink dashes]{HoyasOberlack2022}.}
\end{figure}

A similar phenomenon can be observed in pipe DNS, as demonstrated for example by the indicator function $\Xi$ of \citet{Yao2023} for $\Reytau = 5197$ in figure \ref{refFig5}, shown below in enlarged form as figure \ref{refFig12}. This figure clearly demonstrates a significant change of the slope of $\Xi$ at $Y\approxeq 0.29$, where the grid spacing $\Delta y^+$ reaches a value of around 7.

\begin{figure}
\center
\includegraphics[width=0.65\textwidth]{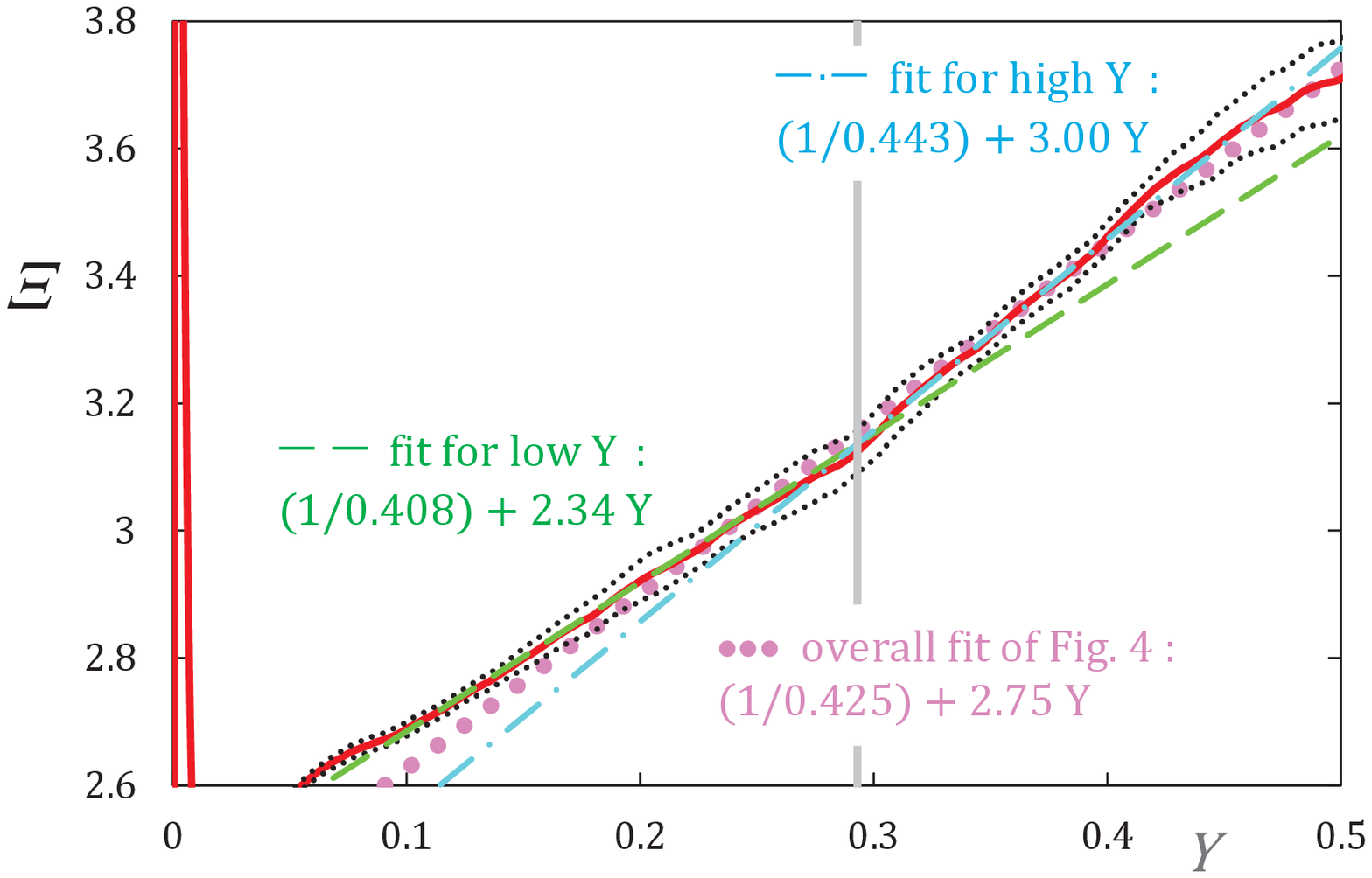}
\caption{\label{refFig12} Detail of $\Xi(Y)$ [(red) ---] for the pipe DNS of \citet[][figure 5]{Yao2023} at $\Reytau = 5197$. (black) $\cdots$, uncertainty estimates given by \citeauthor{Yao2023}; (pink) $\bullet \bullet \bullet$, overall fit $(1/0.425) + 2.75\,Y$ of figure \ref{refFig5}; (green) - - -, best fit $(1/0.408) + 2.34\,Y$ for $Y\in[0.1, 0.29]$; (blue) $-\cdot -$, best fit $(1/0.443) + 3.00\,Y$ for $Y\in[0.26, 0.45]$.}
\end{figure}

As the slope of the linear part $S_0\,Y$ of $\Xi$ is proportional to $\Reytau^{-1}$ in a graph of $\Xi$ versus $y^+$ (see equation \ref{UY}), the steepening of the linear part of $\Xi$ in figures \ref{refFig3} and \ref{refFig12} corresponds to a \textit{decrease} of the effective Reynolds number. This phenomenon is well known from the numerical integration of simpler equations, such as the diffusion equation, where an exaggerated rapid increase of the integration step results in a solution corresponding to a higher diffusivity. This suggests, that the grid spacing in the outer flow contributes significantly to the large differences in figures \ref{refFig3} and \ref{refFig12}, and should probably be limited to $\Delta y^+ \lessapprox 3-4$. However, the fidelity of a DNS to the true Navier-Stokes solution is also influenced by a number of other factors, such as statistical convergence \citep[see for instance][]{Vinetal16}, the order of the numerical scheme, the computational box size, and the ratio of $\Delta \widehat{y}$ to the Kolmogorov length, so that further investigations are clearly called for.\newline

\section{The effect of pressure gradient on the overlap profile - Lucchini's analysis and beyond}\label{appC}

The result of \citet{Luchini17}, that, for sufficiently small pressure gradients, the coefficient $S_0$ of the linear term in the MVP overlap is proportional to the pressure gradient parameter $\beta \equiv -(\widehat{\mathcal{L}}/\widehat{\tau_w}) (\dd \widehat{p}/\dd \widehat{x})$, which equals 1 and 2 for channels and pipes, respectively, can be justified in different ways.

One possibility is to use the stream-wise mean momentum equation (\ref{Umom})
\begin{equation}
\frac{\dd}{\dd y^+}\,\left[\Reytau \nu_T\,\frac{\dd U^+}{\dd y^+}\right] + \frac{\beta}{\Reytau} = 0
\label{Umom}
\end{equation}
with a scaled eddy viscosity $\Reytau \nu_T$. Integrating (\ref{Umom}) once, and switching to the outer coordinate $Y$ yields
\begin{equation}
\nu_T\,\frac{\dd U^+}{\dd Y} = - \beta\,Y - K
\label{Umomint}
\end{equation}
Integrating equation (\ref{Umomint}) once more with the simple (negative) turbulent viscosity \newline $\nu_T = -\ell\,Y$ yields
\begin{equation}
U^+(Y) = \frac{\beta}{\ell}\,Y + \frac{K}{\ell}\,\ln{Y} + C \quad ,
\label{Umomintint}
\end{equation}
where $\ell$, $K$ and $C$ may depend on $\beta$ and are simply related to the overlap part of the mean velocity derivative in equations (\ref{UY}, \ref{UYcofs}) and of the mean velocity in equation (\ref{Uout}). In other words, (\ref{Umomintint}) is perfectly suited to describe all ``canonical'' overlap profiles.

As seen below, it is more general than the overlap derived by \citet{Luchini17}, who used dimensional analysis for the derivation of the $\beta$-dependence of the overlap profile. However, he excluded the channel half-width or pipe radius from the list of variables for the application of the Buckingham $\Pi$ theorem, while implicitly keeping the hydraulic diameter, which is inversely proportional to the pressure gradient. This resulted in a universal $\kappa$, coveted by generations of fluid mechanicians, but not consistent with experimental evidence \citep[see e.g.][]{NagibChauhan2008,Monk17,Monk21}.

Completing Lucchini's list of variables with the channel half-width or the pipe radius $\widehat{\mathcal{L}}$ to $\{\widehat{U}_y , \widehat{y}, \widehat{u}_{\tau}, \widehat{p_x}, \widehat{\mathcal{L}}, \widehat{\rho} \}$, one readily obtains the functional relation $\Pi_1 = f(\Pi_2, \Pi_3)$ between the three non-dimensional $\Pi$'s
\begin{equation}
\Pi_1 = \frac{\widehat{y}\,\widehat{U_y}}{\widehat{u}_{\tau}}~,\quad \Pi_2 = - \widehat{\mathcal{L}}\,\frac{\widehat{p_x}}{\widehat{\tau}_w} \equiv \beta~,\quad \Pi_3 = \frac{\beta\,\widehat{y}}{\widehat{\mathcal{L}}}~.
\label{Pi}
\end{equation}
Linearizing $\Pi_1 = f(\Pi_2, \Pi_3)$ around $\beta = 0$ yields $\Pi_1=\kappa^{-1}+ B\,\Pi_2 + C\,\Pi_3 + \mathcal{O}(\beta^2)$. With $\widehat{\mathcal{L}}\,\widehat{u}_{\tau}\,\widehat{\nu}^{-1}\equiv \Reytau$, this relation integrates to the non-dimensional overlap profile
\begin{equation}
U^+_{\mathrm{overlap}} = \frac{1}{\kappa_0 +\kappa_1\,\beta}\,\ln(y^+) + \left[B_0 + B_1\,\beta\right] + \frac{\beta}{\Reytau}\,\left[C_0 \,y^+ + C_1\right] + \mathcal{O}(\beta^2)
\label{final}
\end{equation}
One readily identifies $\beta\,C_0$ in equation (\ref{final}) with the coefficient $S_0$ in equations (\ref{UY}) and (\ref{UYcofs}).\newline

Declaration of Interests: The authors report no conflict of interest.

\bibliographystyle{jfm}
\bibliography{Turbulence}

\end{document}